\documentclass{aa}  

\usepackage{graphicx}
\usepackage{subcaption}
\usepackage{txfonts}
\usepackage{multirow}
\usepackage{mathrsfs} 
\usepackage{multirow}
\usepackage{booktabs}
\usepackage{threeparttable}

\begin{document}

   \title{Improving pulsar timing precision through superior Time-of-Arrival creation}

   \author{J.~Wang                    \inst{1,2,3}   \fnmsep\thanks{jun.wang.ucas@gmail.com}
        \and    J.~P.~W.~Verbiest     \inst{4}
        \and    G.~M.~Shaifullah      \inst{5,6}
        \and    I.~Cognard            \inst{7,8}
        \and    L.~Guillemot          \inst{7,8} 
        \and    G.~H.~Janssen         \inst{9,10}
        \and    M.~B.~Mickaliger      \inst{11}
        \and    A.~Possenti           \inst{12}
        \and    G.~Theureau           \inst{7,8,13}
   }

    \institute{Ruhr-Universit\"at Bochum, Fakult\"at f\"ur Physik und Astronomie, Astronomisches Institut (AIRUB), 44801 Bochum, Germany
     \and  Advanced Institute of Natural Sciences, Beijing Normal University, Zhuhai 519087, China
     \and  Fakulat\"at f\"ur Physik, Universit\"at Bielefeld, Postfach 100131, 33501 Bielefeld, Germany
     \and Florida Space Institute, University of Central Florida, 12354 Research Parkway, Orlando, 32826, Florida, USA
     \and  Dipartimento di Fisica `G. Occhialini', Universit\`{a} degli Studi di Milano-Bicocca, Piazza della Scienza 3, 20126 Milano, Italy
     \and  INFN, Sezione di Milano-Bicocca, Piazza della Scienza 3, 20126 Milano, Italy
     \and  Laboratoire de Physique et Chimie de l'Environnement et de l'Espace LPC2E CNRS-Universit{\'e} d'Orl{\'e}ans, 45071, Orl{\'e}ans, France;    
     \and  Station de radioastronomie de Nan{\c c}ay, Observatoire de Paris, PSL Research University, CNRS/INSU 18330 Nan{\c c}ay, France 
     \and  ASTRON, the Netherlands Institute for Radio Astronomy, Oude
    Hoogeveensedijk 4, NL-7991 PD Dwingeloo, the Netherlands
    \and Department of Astrophysics/IMAPP, Radboud University Nijmegen, P.O.
    Box 9010, NL-6500 GL Nijmegen, the Netherlands
    \and  Jodrell Bank Centre for Astrophysics, University of Manchester,
     Manchester, M13 9PL, United Kingdom
    \and  INAF – Osservatorio Astronomico di Cagliari, via della Scienza 5, 09047 Selargius (CA), Italy
    \and  Laboratoire Univers et Th{\'e}ories LUTh, Observatoire de Paris, PSL Research University, CNRS/INSU, Universit{\'e} Paris Diderot, 5 place Jules Janssen, 92190 Meudon, France 
   }

   \date{Received MMMMM DD, YYYY; accepted MMMMM DD, YYYY}

   % \abstract{}{}{}{}{} 
   % 5 {} token are mandatory
   
   \abstract
   {The measurement of pulsar pulse times-of-arrival (ToAs) is a crucial step in detecting low-frequency gravitational waves. To determine ToAs, we can use template-matching to compare each observed pulse profile with a standard template. However, using different combinations of templates and template-matching methods (TMMs) without careful consideration may lead to inconsistent results. 
     }
   {In pulsar timing array (PTA) experiments, distinct ToAs from the same observations can be obtained, due to the use of diverse templates and TMMs. In other words, employing diverse approaches can yield different timing results and would thus have a significant impact on subsequent gravitational wave searches. In this paper, we examine several commonly used combinations to analyze their effect on pulse ToAs.}
   {we evaluate the potential impact of template and TMM selection on thirteen typical millisecond pulsars within the European PTA. We employ pulsar timing methods, specifically the root mean square and reduced chi-square $\chi_r^2$ of the residuals of the best timing solution to assess the outcomes. Additionally, we evaluate the system-limited noise floor (SLNF) for each pulsar at various telescopes operating around 1.4~GHz using frequency-resolved templates.}
   {Our findings suggest that utilizing data-derived and smoothed templates in conjunction with the Fourier-domain with Markov-chain Monte Carlo (FDM) TMM is generally the most effective approach, though there may be exceptions that require further attention. Furthermore, we determine that pulse phase jitter noise does not significantly limit the current precision of the European PTA's timing, as jitter levels derived from other studies are much smaller than the SLNF.}
   {}
   

   \maketitle

\section{Introduction} \label{sec:intro}
Since their discovery, millisecond pulsars \citep[MSPs,][]{bkh+82} have been known for their short spin periods and highly stable rotation characteristics. 
These unique properties have allowed for a diverse range of applications in testing fundamental physics. These applications encompass high-precision tests of general relativity \citep[GR,][]{zsd+15, ksm+21}, exploration of the ionized interstellar medium \citep[IISM,][]{tmc+21, lvm+22} and solar wind \citep{tvs+19, tsb+21}, development of pulsar-based timescales \citep{hcm+12, hgc+20}, constraints on the equations-of-state for neutron stars \citep{lp16, of16}, measurements of the model-independent properties of the pulsar itself \citep{vbv+08}, etc.

One of the most significant applications of MSP observations, however, lies in the detection and characterization of low-frequency (nano-Hertz) gravitational waves \citep[GWs, see e.g.][]{vlh+16, pdd+19}. These waves represent fluctuations in the fabric of space-time caused by the propagation of GWs, as predicted by the theory of general relativity \citep[see e.g.,][]{mtw73}. Pulsar timing arrays (PTAs), which consist of an array of MSPs, are particularly sensitive to low-frequency GW signals. In this frequency range, the potential sources of GWs are likely an ensemble of super-massive black-hole binaries, whose combined, stochastic signal gives rise to the GW Background \citep{btc+19}. Typically, several factors can influence the precision of pulsar timing, and each source of noise has its unique impact \citep{vs18}. However, GWs leave a distinct quadrupolar signature in the pulsar timing residuals. A method based on the correlation between pairs of pulsars in the sky, as a function of their angular separation, known as the ``Hellings and Downs'' curve \citep{hd83}, can be employed to identify this signature of GWs.

Pulsar timing analysis involves determining the differences, referred to as timing residuals, between the observed times-of-arrival (ToAs) and the predicted ToAs. These predictions are based on models that take into account the influence of the astrometry and ephemeris of the MSPs in the array, the Earth and Solar system bodies, etc. A precise and comprehensive model would result in accurate timing residuals. However, the precision of the timing, quantified by the root mean square (RMS) value of the timing residuals, is constrained by various sources of noise originating from the pulsar itself, the propagation path, or the observing system. To enhance timing precision and increase the likelihood of detecting GWs, researchers have explored and implemented various analysis methods and hardware upgrades in PTA observations and data analysis. Nevertheless, the impact of different data analysis pipelines at different telescopes has not been thoroughly investigated. To solve this issue, \citet{wsv+22} presented a method to assess the influence of different template options and template-matching Methods (TMMs) using three pulsars. Here we briefly review some of the basic concepts mentioned. The Fourier phase gradient (PGS) determines the shift between two similar profiles by fitting for the gradient in the Fourier domain. The Fourier domain method with Markov-chain Monte Carlo (FDM) is identical to PGS in how it determines ToA values but applies a one-dimensional Markov-chain Monte Carlo method to measure the ToA variance. The Gaussian interpolation shift (GIS) calculates the discrete cross-correlation function between the profile and the template, and a Gaussian curve is then fitted to the resulting points to enable interpolation between each phase bin. For the construction of the single template, the observation with the highest signal-to-noise ratio (S/N) was selected to serve as the template and was excluded from the set of observations to be timed. For the added template, all observations were added up and aligned using the timing ephemeris of the data. For the smoothed template, the added profile was then refined using an undecimated Daubechies wavelet smoothing filter to produce the final smoothed template. The analytic template was generated by fitting von Mises functions to the added profile. For a more detailed description regarding the TMMs and templates, please refer to \citet{wsv+22}.

\citet{wsv+22} concluded that data-derived and smoothed templates are typically preferable, and that the FDM method generally outperforms other methods. However, given the limited number of pulsars in \citet{wsv+22}, these findings should be examined and verified with a larger sample of pulsars. More recently, \citet{wvsy23} delved into TMMs for pulsars, scrutinizing the shape of the pulse profile, and exploring the impact of the S/N of both the template and the observation on the ToA uncertainty. They achieved this by examining pulsar pulse profiles obtained from the International Pulsar Timing Array (IPTA). Through an extensive analysis involving numerous simulation models and real data, \citet{wvsy23} discovered that while the ToA uncertainty computed with the FDM method is generally more reliable than other methods, there are instances where alternative methods prove to yield more well-behaved ToA uncertainties compared to FDM. This occurs particularly in cases where the duty cycle of the pulse profile is large, the S/N of the observations is low, or the pulse profile lacks sharp features. However, since the study only focused on ToA uncertainties, the optimal method for determining both ToAs and their uncertainties requires further investigations.

In this study, we aim to elaborate on the findings reported in \citet{wsv+22} and \citet{wvsy23} using thirteen MSPs observed with European telescopes. The selection of these pulsars is based on their significance within the EPTA, as determined by \citet{bps+16}. Consequently, we scrutinize the achieved timing precision as a function of the chosen templates and TMMs. Additionally, we assess the noise floor for each pulsar across the four EPTA telescopes. The remainder of this article is structured in the following manner. In Section~\ref{sec:observation}, we provide an overview of the observations and the methodologies employed for data reduction in this study. The timing results are presented on a per-pulsar basis in Section~\ref{sec:disscuss}. Finally, we discuss these results and summarize our conclusions in Section~\ref{sec:conclusion}.

\section{Observations and Analysis Procedure} \label{sec:observation}
This paper presents observations of an array of MSPs at L-band\footnote{L-band refers to the frequency range of 1–2~GHz in the radio spectrum.} from four EPTA telescopes over 13 years, from 2007 to 2020. The EPTA is a collaborative effort among European countries with the primary goal of directly detecting low-frequency GWs. The concept and objectives of this organization were initially outlined in detail by \citet{skl+06}, and a more comprehensive overview of the EPTA telescopes, the observing system, and the list of primary sources can be found in the work of \citet{eab+23}.

The data set discussed in this paper was collected by four European telescopes: the Effelsberg 100-m radio telescope (EFF) in Germany, the Lovell radio telescope at the Jodrell Bank Observatory (JBO) in the UK, the Westerbork Synthesis Radio Telescope (WSRT) in the Netherlands, and the Nan{\c c}ay decimetric radio telescope (NCY) in France. At each telescope, pulsar observations were carried out using the latest generation of instruments, namely the PSRIX \citep{lkg+16}, ROACH \citep{bjk+16}, NUPPI \citep{ctg+13,ldc+14}, and PuMa II \citep{ksv08}. All thirteen sources analyzed in this study were observed at intervals ranging from weekly to monthly as part of the EPTA program, except for NCY which was observed more frequently. At EFF, observations were recorded across a bandwidth of 200~MHz with a center frequency of 1347.5~MHz. Data collected from both the multi-beam and single-feed 21~cm receivers were utilized for this study. Typically, each pulsar was observed once a month with an integration time of approximately 30~minutes. It should be noted that due to the scheduling of data acquisition, the calibration files were missing when the EFF observation data were collected. As a result, the EFF data were not polarization-calibrated. Consequently, the analysis results of the EFF data should not be interpreted as indicating any inherent disadvantages compared to data from other telescopes. Pulsars observed with the Lovell telescope at JBO were recorded using the ROACH back-end, with a center frequency of 1532~MHz and a total bandwidth of 400~MHz. NCY observations spanned approximately 10 years, from 2011 to 2020, with an average cadence of about 5 to 10 days. Data were recorded at NCY with a center frequency of 1484~MHz. The total bandwidth of 512~MHz was channelized into 128 channels and dedispersed coherently. Recently, \citet{gcv+23} have deployed an improved polarization calibration scheme for the Nan{\c c}ay NUPPI observations. However, the data presented in our work were not calibrated with the improved scheme, thus the profiles in our data may display variations from (time-varying) imperfect calibration parameters. WSRT observations were conducted at a center frequency of 1380~MHz across a bandwidth of 160~MHz. At WSRT, each pulsar was observed for approximately 30~minutes at intervals of around one month. Further details regarding the bandwidths, frequencies, and average lengths of the observations at each telescope are provided in Table~\ref{table:tel_parameter}.

\begin{table}
    \caption[]{Information regarding the observational configurations of various EPTA telescopes pertinent to this study.}
    \centering
        \scalebox{0.9}{
\begin{tabular}{ccccc}
    \hline
    Telescope             &   EFF     &    JBO    &    NCY    &   WSRT  \\
    \hline
    Pulsar back-end       &  PSRIX    &   ROACH   &   NUPPI   &   PuMa II  \\
    Frequency (MHz)       &  1347.5   &   1532    &   1484    &   1380     \\
    Bandwidth (MHz)       &  200      &   400     &   512     &   160      \\
    Average length (min)  &  30       &   20      &   50      &   30       \\
    Number of channels    &  128      &   400     &   128     &   512      \\
    Number of phase bins  &  1024     &   512     &   2048    &   512      \\
    \hline
\label{table:tel_parameter}
\end{tabular}}
\end{table}

In Tables~\ref{tab:peo} and \ref{tab:pep}, we provide a summary of the pulsar details, the Modified Julian Date (MJD) ranges, and the number of observations retained after the initial data quality check.

The timing models utilized in this study are primarily based on the work of \citet{dcl+16}, with updates to the reference clock and the Solar system ephemeris to ``TT(BIPM2019)'' and ``DE438'',  respectively. Since our focus is not on investigating the parameters of the timing model, we only fit the rotational frequency and its first derivative in the analysis of templates and TMMs. However, due to variations in the dispersion measure (DM) at different sub-bands, we also fitted the DM, along with its first and second derivatives whenever possible, for the investigation of the system-limited noise floor (SLNF), which refers to the level of residual noise saturation observed in pulsar timing measurements when using the widest ToA bandwidths.

Prior to any further processing, it is essential to remove any radio-frequency interference (RFI) from the data. Initially, all data, except for those from JBO, underwent processing using the \texttt{clean.py} tool from the \textsc{CoastGuard} package \citep{lkg+16} to eliminate contaminated channel-subintegration combinations. Any remaining RFI was then manually and interactively removed through visual inspection using the \textsc{PSRCHIVE} \citep{hvm04} \texttt{pazi/paz} tools. For JBO data, a specific Python package called \textsc{CLFD} \citep{mbc+19} was automatically applied. This package identifies and zero-weights RFI using Tukey's rule \citep{tuk77} in a three-dimensional feature space (standard deviation, peak-to-peak difference, and absolute value of the second bin of the profile's Fourier transform) before the data are manually inspected.

\renewcommand{\arraystretch}{1.5} 
\begin{table*}
    \caption{Overview of the pulsar observations. The range of Modified Julian Dates (MJDs) for pulsars observed by different telescopes, along with the number of obtained observations $(N_{obs})$ and the overall data span, is presented for each pulsar-telescope combination.}
  \centering
  \fontsize{6.5}{8}\selectfont
  \begin{threeparttable}

    \begin{tabular}{cccccccccc}
    \toprule
    \multirow{2}{*}{PSR} & \multicolumn{2}{c}{EFF} & \multicolumn{2}{c}{JBO}&\multicolumn{2}{c}{NCY}&\multicolumn{2}{c}{WSRT}  & \multicolumn{1}{c}{Data span} \cr
    \cmidrule(lr){2-3} \cmidrule(lr){4-5} \cmidrule(lr){6-7} \cmidrule(lr){8-9}
    & MJD Range  &  $N_{obs}$  &  MJD Range  &  $N_{obs}$  &  MJD Range &  $N_{obs}$  &  MJD Range  &  $N_{obs}$  & (yr) \cr
    \midrule
     J0030+0451    &  55633-58859  & 55  & 56095-58495 & 101 & 55803-59152 & 706 & 56080-57113 & 35   &  9.6   \cr
     J0613$-$0200  &  55600-58620  & 91  & 56253-58494 & 83  & 55817-58851 & 318 & 54337-57096 & 101  &  12.4  \cr
     J1012+5307    &  55633-59083  & 103 & 55655-58495 & 144 & 55816-59288 & 799 & 54165-57153 & 83   &  14.0  \cr
     J1022+1001    &  55660-58769  & 109 & 55600-58131 & 184 & 55839-58853 & 280 & 54154-57153 & 80   &  12.9  \cr
     J1024$-$0719  &  55633-59223  & 79  & 55674-59120 & 158 & 55891-59153 & 393 & 54287-57118 & 75   &  13.3  \cr
     J1600$-$3053  &  55661-58621  & 60  & 56255-58559 & 84  & 55809-58852 & 624 & 54155-57131 & 62   &  12.9  \cr
     J1640+2224    &  55633-58620  & 94  & 55653-58467 & 95  & 55826-58850 & 369 & 54155-57101 & 86   &  12.9  \cr
     J1744$-$1134  &  55628-59161  & 87  & 55670-58560 & 94  & 55805-59212 & 213 & 55792-57154 & 52   &   9.4  \cr
     J1857+0943    &  55633-58824  & 64  & 56011-58559 & 121 & 55800-58837 & 235 & 54337-57177 & 81   &   12.4 \cr
     J1909$-$3744  &       -       & -   &      -      &  -  & 55812-59277 & 515 &      -      & -    &   9.5  \cr
     J1918$-$0642  &  55633-59223  & 76  & 55652-58838 & 86  & 55874-59138 & 169 & 54490-57190 & 86   &   13.0 \cr
     J2010$-$1323  &  56206-59223  & 34  & 55657-58494 & 96  & 55850-59149 & 298 & 54492-57148 & 75   &   13.0 \cr
     J2317+1439    &  55633-58824  & 58  & 55662-58495 & 83  & 55850-58849 & 701 & 54490-57117 & 78   &   11.9 \cr
    \bottomrule
    \end{tabular}
    \end{threeparttable}
    \label{tab:peo}
\end{table*}

\renewcommand{\arraystretch}{1.5} 
\begin{table*}
    \caption{Overview of the pulsars' properties. The pulsar's rotation period $P$, spindown $\dot{P}$, orbital period $P_{\rm b}$, DM, and flux density at 1.4 GHz, $S_{1.4}$, are outlined.}
  \centering
  \fontsize{6.5}{8}\selectfont
  \begin{threeparttable}

    \begin{tabular}{ccccccc}
    \toprule
    PSR & $P$  (ms) & $\dot{P} (10^{-20}$) & $P_{\rm b}$  (day) &  DM (cm$^{-3}$ pc) & $S_{1.4}$ (mJy) & Ref.  \cr
    \midrule
     J0030+0451    & 4.865  &  1.02   &   -     &   4.33  &  0.6  &  \citet{lzb+00, vlh+16}                \cr
     J0613$-$0200  & 3.062  &  0.96   &  1.20   &  38.79  &  1.7  &  \citet{lnl+95, vlh+16}                \cr
     J1012+5307    & 5.26   &  1.71   &  0.60   &  9.02   &  2.8  &  \citet{nll+95, lcw+01}                \cr 
     J1022+1001    & 16.453 &  4.33   &  7.81   &  10.25  &  6.1  &  \citet{cnst96, dcl+16}                \cr
     J1024$-$0719  & 5.162  &  1.86   &   -     &  6.49   &  1.5  &  \citet{bjb+97, mhb+13}                \cr 
     J1600$-$3053  & 3.598  &  0.95   &  14.35  &  52.33  &  2.5  &  \citet{jbo+07,ojhb06, mhb+13}         \cr 
     J1640+2224    & 3.163  &  0.28   &  175.46 &  18.43  &  0.4  &  \citet{fcwa95, dcl+16}                \cr 
     J1744$-$1134  & 4.075  &  0.86   &  -      &  3.14   &  1.0  &  \citet{bjb+97}                        \cr
     J1857+0943    & 5.362  &  1.78   &  12.33  &  13.30  &  3.3  &  \citet{srs+86, vlh+16}                \cr
     J1909$-$3744  & 2.95   &  1.40   &  1.53   &  10.39  &  1.0  &  \citet{jbv+03, dcl+16}                \cr   
     J1918$-$0642  & 7.646  &  2.57   &  10.91  &  26.55  &  0.6  &  \citet{eb01b, jsb+10}                 \cr
     J2010$-$1323  & 5.223  &  0.48   &   -     &  22.16  &  1.6  &  \citet{jbo+07}                        \cr 
     J2317+1439    & 3.445  &  0.24   &  2.46   &  21.90  &  4.0  &  \citet{cnt93, vlh+16}                 \cr
    \bottomrule
    \end{tabular}
    \end{threeparttable}
    \label{tab:pep}
\end{table*}

\section{Results}  \label{sec:disscuss}

In this section, we present the specific findings for each of the 13 pulsars in our sample and evaluate how the choice of template and TMM affects data from different EPTA telescopes. In addition, we apply the SLNF analysis and determine the optimal ToA bandwidth. We begin by discussing individual pulsars with distinctive results. For pulsars that exhibit similar outcomes, we provide a collective description at the end of this section.

Given that we have previously highlighted the subpar performance of templates from the single brightest observation and of the GIS TMM \citep[see][]{wsv+22}, and because we consistently observe the same pattern from the majority of our current sample, we will refrain from reiterating the comparison and description of the single template and GIS TMM analysis in the subsequent text, unless a specific and unique result emerges for a particular pulsar, telescope, and back-end combination.

\subsection{PSRs~J0030+0451 and J2010$-$1323} \label{ssec:0030_2010}
PSRs~J0030+0451 and J2010$-$1323 are two of the four isolated MSPs in our sample. PSR~J0030+0451 is one of the closest known MSPs in the Galaxy, situated at a distance of $\sim$325 parsecs \citep{lzb+00, abb+18}. Due to the likelihood of strong influence from solar wind at low solar elongations, observations taken when the angular separation from the Sun is less than 5 degrees \citep[see][for detailed discussions]{tsb+21, vlh+16} have been excluded from our analysis. PSR~J2010$-$1323 was discovered by \citet{jbo+07} during the high Galactic-latitude survey. 
% Its pulsed emission has been detected across various electromagnetic frequencies, including X-ray \citep{btlb00}, radio \citep{lzb+00}, and $\gamma$-ray \citep{aaa+09e}.

At some of the EPTA telescopes, only a small number of observations were made for these two pulsars, and thus our results must be interpreted with caution for those datasets. PSR~J0030+0451 was observed only 35 times at WSRT, while PSR~J2010$-$1323 had 34 observations with EFF. In addition, with mean S/N values of 25.30 and 50.02, respectively, both of these pulsars have low S/N. The timing analysis and SLNF fitting are significantly impacted by the small sample size and low S/Ns. The outlier rejection strategy, as defined in \citet{wsv+22}, which primarily relies on the S/N criterion, exacerbates these issues, resulting in an even smaller sample size.

In the cases of EFF and WSRT, the results of the template and TMM choices lead to two completely opposing outcomes. For PSR~J0030+0451 at WSRT, while most other combinations yield consistent and intuitive results, the added template combined with PGS results in an almost halved RMS of the residuals compared to FDM. The superiority of the added template in this case is likely due to imperfections in both the smoothing algorithms and von Mises functions fitting process, which remove certain features (as can be seen in Fig~\ref{fig:pprof1}) in the added template. In other words, the smoothed and analytic templates do not always accurately represent the true profile. However, the situation for PSR J2010$-$1323 at EFF is completely reversed, with the added template combined with PGS resulting in an almost doubled RMS compared to FDM. Both combinations are outperformed by other options. After inspecting the profile differences, no distinct features were found between the added template and the analytic template, indicating that the noise-free template can accurately represent the profile. Therefore, we attribute the improvement to the removal of random noise.

Regarding the SLNF analysis, the fitting for these two special cases fails due to the limited number of ToAs and low S/N. In cases with small ToA bandwidths, a larger proportion of ToAs are eliminated by our outlier rejection scheme, leading to biased RMS values. Aside from the specific cases mentioned above, when these two pulsars are observed with other telescopes, the results align with the conclusions drawn in \citet{wsv+22}: added and smoothed templates are typically preferred over some more commonly applied alternatives, and the FDM method is generally superior to or competitive with other algorithms. At NCY, PSR~J0030+0451 is the most frequently observed pulsar in our dataset, with a cadence of 5 days. The RMS values of PSR~J0030+0451 obtained with NCY decrease more rapidly than with other telescopes as a function of ToA bandwidth, and the asymptotic RMS limit, as low as 1.18 $\mu s$, appears to be reached with a ToA bandwidth of about 1~GHz. With an SLNF of 1.45 $\mu s$, which is achievable with the existing backend, NCY is likewise the best-performing telescope for PSR~J2010$-$1323. Extending the available bandwidth is not expected to significantly increase timing precision.

\subsection{PSR~J1012+5307}
PSR~J1012+5307, discovered by \citet{nll+95}, resides in a 14.5-hour binary system with an extremely low-mass helium-core white dwarf companion, estimated to be around $0.16 M_{\odot}$ \citep{vbk96, cgk98}. Leveraging data from EFF and JBO, \citet{lcw+01} established a precise timing model for PSR~J1012+5307, and determined the detailed 3D velocity motion for this binary system through a combination of radio timing and optical observations. A subsequent study by \citet{lwj+09} revisited PSR~J1012+5307, further refining the 3D velocity information with 15 years of EPTA data. One noteworthy outcome was the establishment of a new upper limit for intrinsic eccentricity, which is less than $8.4 \times 10^{-7}$, positioning it as one of the smallest values recorded for a binary system.

\begin{figure}
\centering
 \includegraphics[width=0.9\columnwidth]{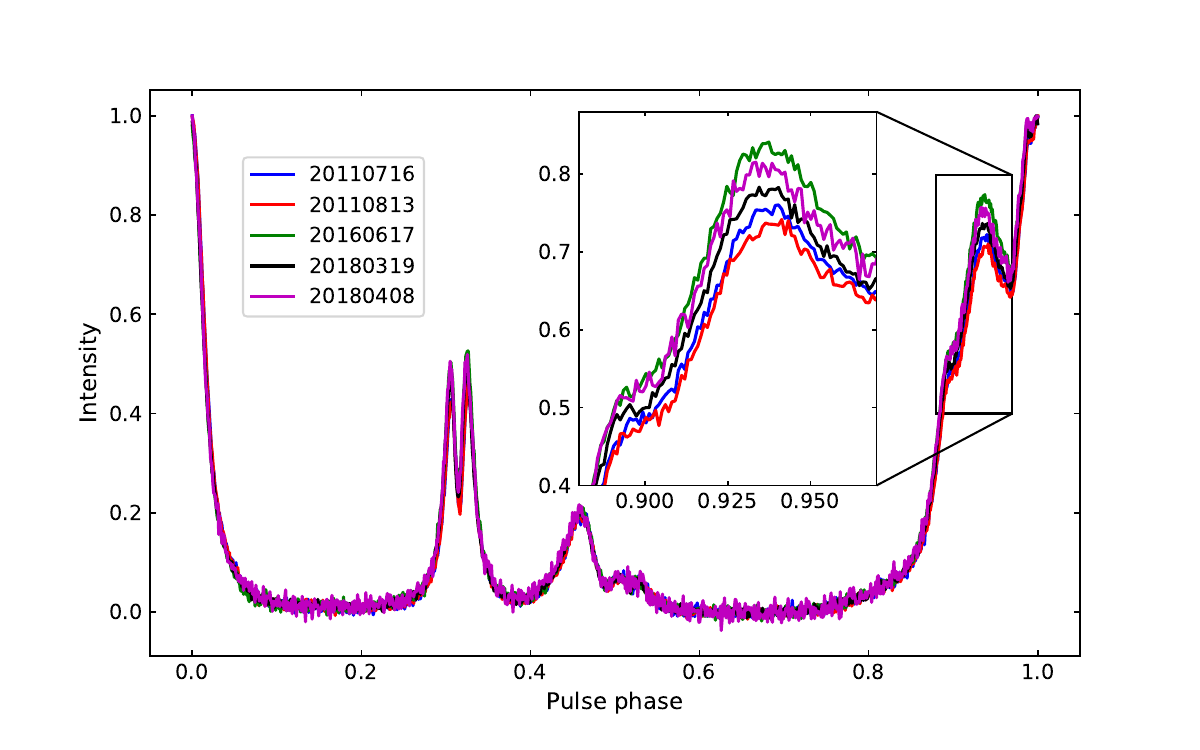}
 \caption[Profile variations of PSR~J1012+5307 at EFF.]{Example of profile variations of PSR~J1012+5307 at EFF, with each color representing a peak-normalized observation taken at a certain date, as indicated in the legend. The leading component of the main pulse is zoomed in at the subplot. These variations are not found in other telescopes, and thus we attribute these variations to the fact that we did not polarization calibrate the EFF data.}
 \label{fig:j1012_eff_profdiff}
\end{figure}

PSR~J1012+5307 is among the most frequently observed, as well as having one of the longest data-spans pulsars within the EPTA, particularly with the NCY telescope. While this pulsar is observed approximately once per month at EFF, JBO, and WSRT, NCY observes this pulsar roughly every six days. In the case of EFF data, this pulsar exhibits a significantly high reduced chi-square $\chi_r^2$, likely attributable to profile variations stemming from uncalibrated polarization (refer to Figure~\ref{fig:j1012_eff_profdiff}). Except for this telescope, the conclusions drawn are consistent. The combination of the added template and FDM method produces more reliable and precise ToAs. PGS may yield a slightly improved RMS value, but it substantially elevates the $\chi_r^2$. The SLNF fitting results depicted in Figure~\ref{fig:psr_toabw} indicate that this pulsar's performance observed with EFF is suboptimal, and increasing the ToA bandwidth does not significantly reduce the RMS. This is most likely due to our failure to calibrate the EFF data.

\subsection{PSR~J1022+1001}
\begin{figure}[htbp]
 \includegraphics[width=0.9\columnwidth]{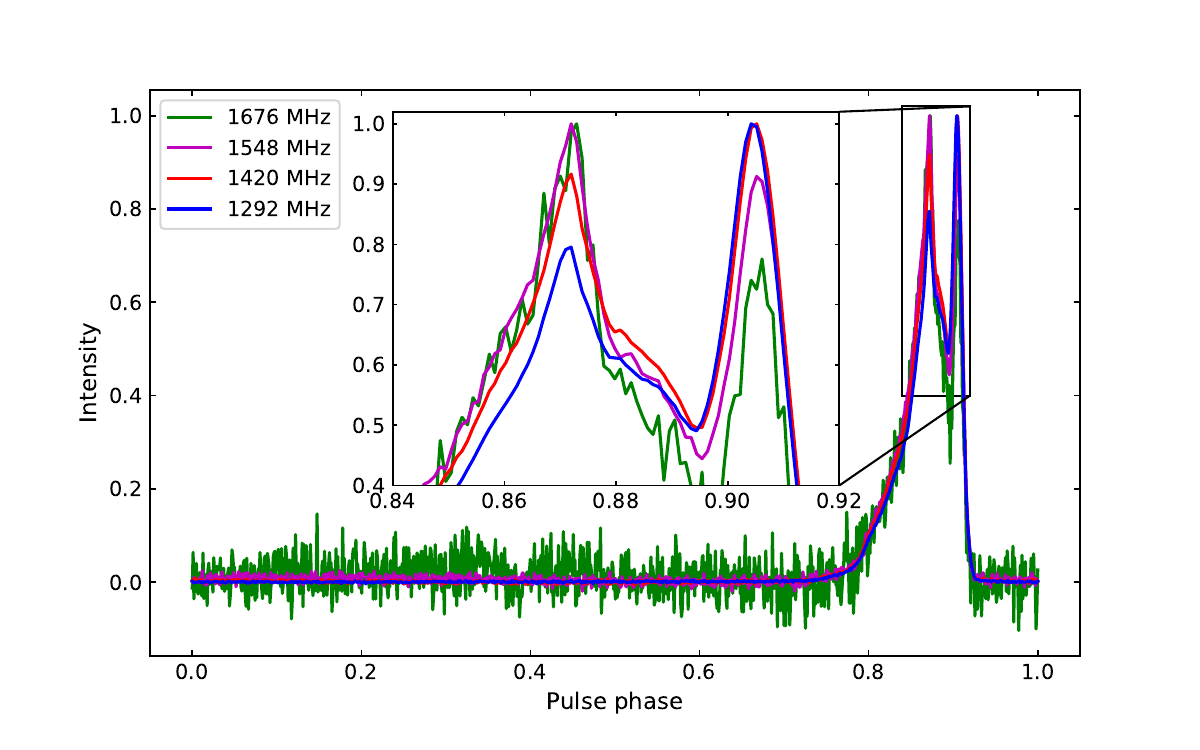}
 \caption{An example of profile variation of PSR~J1022+1001 at NCY as a function of frequency. The entire band is divided into 4 subbands with a subband width of 128 MHz. The data underwent polarization calibration and each subband profile was peak-normalized.}
 \label{fig:j1022_ncy_profdiff}
\end{figure}

PSR~J1022+1001 is an MSP with a spin period of 16.45~ms, discovered concurrently by the Princeton-Arecibo Declination-Strip Survey \citep{cnst96} and the Green Bank Northern Sky Survey \citep{snt97}. It is in a binary system with a 7.8-day orbit around a CO white dwarf companion.
The average pulse profile of PSR~J1022+1001 at L-band displays two prominent peaks. While the trailing peak component exhibits high linear polarization, the leading component does not show strong polarization \citep{hbo04}. The pulse profile of PSR~J1022+1001 is observed to undergo significant intrinsic variations with frequency \citep[e.g.\ see Figure~\ref{fig:j1022_ncy_profdiff}, and also][]{rk03, lkl+15}.  Due to strong scintillation at L-band, the intensity ratio of its leading and trailing peaks varies noticeably over time. While \citet{kxc+99} argue that the profile variations in the integrated pulse profile cannot be attributed to instrumental and propagation effects, \citet{van13} find that instrumental polarization artifacts play a major role in the profile instability. \citet{van13} propose a new polarimetric calibration method, referred to as measurement equation template matching (METM), to address this instability. Based on this novel method, timing results from 7.2 years of data of this pulsar from the Parkes radio telescope show that the standard deviation of arrival time residuals derived from METM-calibrated data is two times smaller than with conventional calibration methods.

In addition to profile variations, this pulsar is also located at a very low ecliptic latitude (as low as $-0.06$ deg), suggesting that the pulsar signals could be significantly influenced by solar wind. Generally, dispersion in the solar wind leads to DM fluctuations and an incomplete model of those variations can introduce delays of up to a few microseconds in the ToAs \citep{tsb+21}. As only a small fraction of the observations were taken within five degrees of the Sun, we excluded the observations taken when the Solar elongation was less than five degrees.

Within our dataset, we observed significant profile variations across the observed frequency band, especially in EFF, JBO, and NCY data. In WSRT data, the profile variations are still noticeable but not as pronounced as in other telescopes (presumably due to the narrower bandwidth). Due to these profile variations, ToAs obtained with these telescopes exhibit high $\chi_r^2$-values. However, despite the issue of high $\chi_r^2$-values,  the timing precision at all telescopes remains consistent with each other. 
% This can be expected due to the extremely strong radiation of this pulsar.

For this pulsar, the WSRT data exhibits the lowest SLNF at approximately 1.22 $\mu s$. Additionally, the RMS decreases gradually as a function of ToA bandwidth, similar to JBO and NCY. However, the EFF data behave differently. The increase in ToA bandwidth does not lead to as significant an improvement in RMS as observed in the other datasets, which might be attributed to the lack of polarization calibration for those data.

\subsection{PSR~J1600$-$3053}
PSR J1600$-$3053 is relatively bright for an MSP, especially considering its large DM value as shown in Table~\ref{tab:pep}. Discovered in the Swinburne high Galactic-latitude pulsar survey by \citet{jbo+07}, this binary MSP has since been well-timed with very high precision, owing to its inherently narrow pulse profile and high brightness.

\begin{figure}[ht]
 \includegraphics[width=0.9\columnwidth]{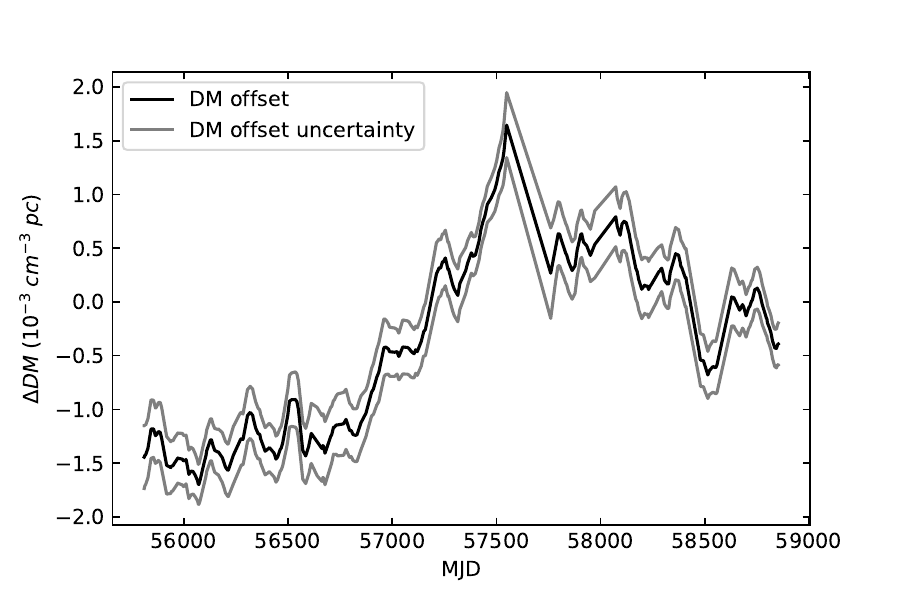}
 \caption[DM time series for PSR~J1600$-$3053]{DM time series for PSR~J1600$-$3053 that was obtained with NCY data, with a reference DM of 52.321 $\rm{cm}^{-3}\; \rm{pc}$.}
 \label{fig:J1600_dmvar}
\end{figure}

During our analysis, we observed that this pulsar is affected by strong DM variations from the end of 2014 onward. Figure~\ref{fig:J1600_dmvar} displays the DM time series for PSR~J1600$-$3053, determined with the NCY dataset. We calculated the DM offset using the method described in \citet{dvt+20} and introduced the `-dmo' flag in the ToA files to correct for these variations. After correcting for the effects of DM variations, the results for templates and TMMs are shown in Figure~\ref{fig:temp_diff01}. We observed that for this pulsar, the added template works very effectively. The most significant improvement is seen in the WSRT data, where the combination of the added template with the PGS method leads to a substantial decrease in RMS and brings the $\chi_r^2$ value in line with the data from other telescopes.

For this pulsar, the profile difference (see Figure~\ref{fig:pprof1}), between the added and smoothed, analytic templates, is noticeable, particularly with the NCY data. These characteristics were not alleviated by employing other smoothing algorithms or fitting with additional Gaussian functions, which could explain the modest deterioration in the performance of the smoothed and analytic templates.

Regarding the SLNF, it is evident that the RMS decreases noticeably as the ToA bandwidth increases. JBO and NCY exhibit a similar rate of decrease, although their final SLNF values differ significantly, being approximately 566~ns and 900~ns, respectively. Another noteworthy observation is that the SLNF fit for WSRT data failed with this pulsar. We attribute this to the low S/N, with a mean S/N of around 20.46 for the observations. Low S/N levels in observations could lead to underestimated ToA uncertainties and result in extra outliers, particularly in frequency-resolved timing where the S/N is diminished with a decrease in ToA bandwidth. Moreover, low S/N levels can lead to a significant number of ToAs being removed by the outlier rejection scheme. For a ToA bandwidth of 20~MHz or less, over 40\% of measurements can be eliminated for this pulsar.

\subsection{PSR~J1909$-$3744}
PSR~J1909$-$3744 is the fastest-spinning MSP in our dataset, with a spin period of 2.95~ms, orbiting in a 1.53-day binary system. It is prominent for its extremely stable pulse profile characterized by a narrow peak and a small pulse duty cycle of approximately 1.5\%. Due to its low declination, in the EPTA it is exclusively observed by the NCY telescope. Using data from NCY, \citet{lgi+20} managed to achieve a timing precision of approximately 100~ns on a 15-year time scale. Notably, studies by \citet{sod+14} and \citet{ls15} have revealed that this pulsar exhibits an unexpectedly low level of jitter noise (around 10~ns), indicating significant potential for further improvements in timing precision.

In light of the presence of achromatic red noise and DM noise components \citep{lgi+20, cbp+22}, the initial RMS and $\chi_r^2$ values are relatively high, standing at 0.241 ms and approximately 25, respectively. To address this, we applied the red noise model established in \citet{lgi+20}, resulting in substantial improvements in the timing solution. The RMS and $\chi_r^2$ dropped significantly to approximately 70~ns and 1.8, respectively.

\begin{figure}[htbp]
 \includegraphics[width=0.9\columnwidth]{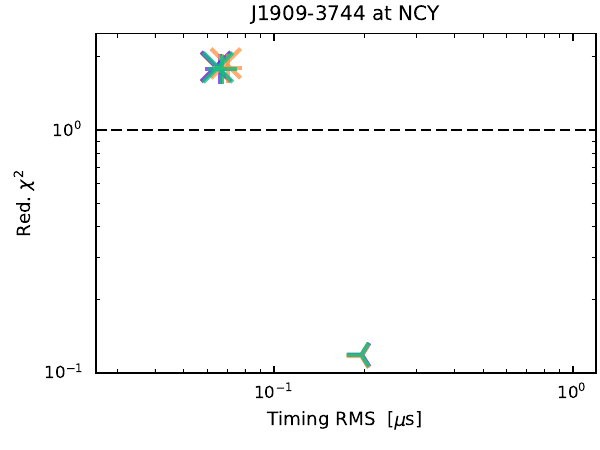}
 \caption[$\chi_r^2$ and residual RMS for PSR~J1909$-$3744 at NCY.]{$\chi_r^2$ and residual RMS for PSR J1909$-$3744 at NCY as a function of the selected TMM and template. In the plot, the "+", "x", and "tri-left" symbols represent the three TMMs, FDM, PGS, and GIS, respectively. The different colors represent the different templates: added (blue), analytic (cyan), single ( orange), and smoothed (olive). As seen in this plot, the GIS technique, compared with other techniques, produces much larger RMS values and inaccurate ToA error estimates.}
 \label{fig:j1909_rmschi}
\end{figure}

\begin{figure}[ht]
 \includegraphics[width=0.9\columnwidth]{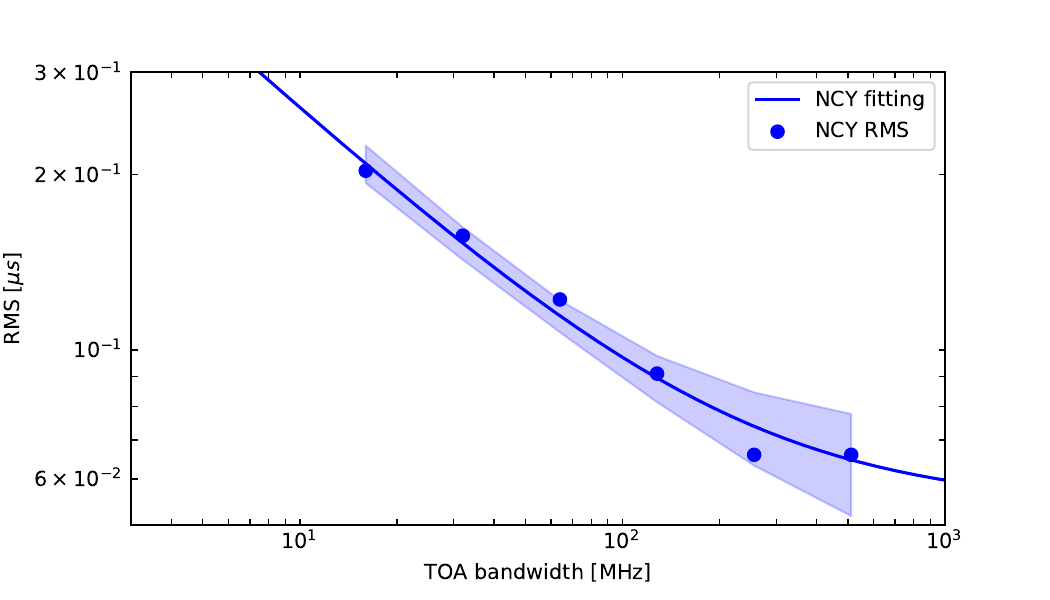}
 \caption{SLNF estimates for PSR~J1909$-$3744 when the added template with the FDM TMM is used to generate the ToA with the available bandwidth at NCY divided into the respective number of channels. The shaded area along the fitted curve is the interquartile range for the fit, representing the error bound on the estimated SLNF.}
 \label{fig:j1909_toabw}
\end{figure}

Figure~\ref{fig:j1909_rmschi} illustrates that for this exceptionally precisely timed pulsar, we observed no notable differences after exploring various combinations of templates and TMMs, when we excluded the obviously inaccurate behavior of GIS. Additionally, we found that the discrepancy between the mean error bars was negligible. In terms of the ToA bandwidth analysis, the noise fitting curve, as depicted in Figure~\ref{fig:j1909_toabw}, aligns with the RMS trend as a function of ToA bandwidth remarkably well. The fitting also suggests that solely increasing the ToA bandwidth of the NCY backend would not result in a significant enhancement in timing precision. We estimated the SLNF for this pulsar to be approximately 54~ns.

\subsection{PSR~J1918$-$0642}
PSR~J1918$-$0642 is an MSP with a spin period of 7.65 ms, located in a binary system with an orbital period of 10.91 days. It was initially discovered by \citet{eb01} in a survey of intermediate Galactic latitudes. \citet{jsb+10} conducted a detailed study of this pulsar using 7.5 years of EPTA observations and updated its timing model. They observed significant flux density modulation in PSR~J1918$-$0642 due to refractive and diffractive scintillation. Subsequent studies by \citet{dcl+16} and \citet{fpe+16} using 12.5 years of EPTA data and 9 years of NANOGrav data, respectively, successfully detected the Shapiro delay in this system.

In our analysis of PSR~J1918$-$0642, we found that the results obtained from JBO data align with the conclusions of \citet{wsv+22}. However, for EFF, NCY, and WSRT ToAs, we observed that using added templates with the FDM method resulted in approximately 10\% higher RMS compared to using PGS with added templates. This difference was most pronounced at WSRT, where PGS led to a 15\% improvement in both RMS and $\chi_r^2$.

\subsection{Other Pulsars} \label{ssec:other_pulsars}
The results for the remaining six pulsars in our dataset are consistent with the conclusions drawn in \citet{wsv+22}. The FDM method consistently outperforms or is on par with other TMMs. Additionally, the added, smoothed, and analytic templates yield comparable results. In certain cases, such as PSR~J0613$-$0200, using the PGS method with added templates results in a slightly lower RMS, but at the expense of increased $\chi_r^2$. This suggests that for PSR~J0613$-$0200, PGS tends to underestimate the ToA uncertainty. Furthermore, PSRs~J1640+2224 and J2317+1439 demonstrate that using the added template in combination with PGS leads to suboptimal determination of ToAs and their uncertainties, as evidenced by noticeably higher RMS and $\chi_r^2$ values.

For PSRs~J1024$-$0719, J1744$-$1134, and J1857$-$0642, the results are consistent across most combinations of templates and TMMs yield similar outcomes. We also observed that ToAs determined by FDM and PGS are the most reliable in terms of ToA uncertainties. FDM tends to perform significantly better than PGS, particularly when templates are not noise-free, resulting in lower $\chi_r^2$ values.

Regarding the SLNF, different pulsars and telescopes yield markedly different results. With a few exceptions, such as PSR~J1744$-$1134, where EFF and WSRT achieve the best SLNF at approximately 300~ns, NCY consistently attains the lowest SLNF. Additionally, EFF exhibits the sharpest decreasing trend, suggesting that adopting a suitable ToA bandwidth can help reduce system noise and bring it closer to the SLNF limit.

\section{Conclusions} \label{sec:conclusion}
This study focuses on evaluating the impact of template and TMM choices on pulsar timing precision for a large sample of EPTA pulsars. Our findings demonstrate that the GIS method is not suitable for achieving high-precision timing. Regardless of the template used, GIS consistently resulted in relatively high levels of RMS values for the timing residuals. Additionally, it was observed that GIS tended to overestimate the ToA uncertainties, leading to seemingly over-fitted timing residuals. Moreover, FDM generally outperforms the PGS algorithm, with a few exceptions in our dataset, typically due to a limited number of ToAs and low-S/N of observations (e.g., PSR~J0030+0451 at WSRT and PSR~J2010$-$0719 at EFF) or underestimated ToA uncertainties (e.g., PSRs~J1640+2224 and J2317+1439). Based on our results, we recommend using FDM as the default method for high-precision pulsar timing analysis.

In terms of templates, we highly recommend using data-derived and smoothed templates. While these templates may be susceptible to the `self-standarding' issue discussed in \citet{hbo05a}, we did not observe any evidence for this in our templates. It is, in this regard, worth noting that the `self-standarding' problem tends to diminish when the S/N of individual observations is above 25 \citep{hbo05a}. Given that most of our observations have a high S/N, this issue has little impact on our investigation. Furthermore, with ongoing hardware upgrades and the development of RFI mitigation algorithms, `self-standarding' is becoming less of a concern for PTA analysis. Therefore, we recommend using added templates as the default for current and future PTA ToA determination.

When categorizing templates and TMMs by telescope, several conclusions become evident. Firstly, all templates and TMMs yield stable results with NCY data, with negligible differences in RMS or $\chi_r^2$ values. This suggests that the choice of template or TMM has minimal impact on NCY data. JBO data shows a similar pattern to NCY, except that PGS combined with added templates can sometimes lead to unsatisfactory results, particularly for PSRs~J0030+0451, J1640+2224, and J2317+1439, which tend to exhibit significantly high $\chi_r^2$ values. With WSRT data, in contrast to the added template, analytic and smoothed templates yield consistent timing results. While PGS combined with added templates may result in high $\chi_r^2$ values for most pulsars, PSRs~J0030+0451 and J1600$-$3053 show significant RMS improvements with the added template. For EFF, the added template combined with FDM is generally preferred, as it tends to yield either better RMS and $\chi_r^2$ values or consistent results. However, there are two exceptions, PSRs~J1744$-$1134 and J2010$-$1323, where smoothed and analytic templates provide consistent timing results, both superior to the added template.

Overall, our findings suggest that with the improvement of telescope sensitivity, template and TMM choices will have a smaller impact on pulsars currently timed for PTA purposes. However, for the time being, it is important to carefully consider TMM and template selection, especially for telescopes with lower sensitivity. Based on available data and experimental comparisons, we strongly recommend using a combination of data-derived and smoothed templates, along with the FDM method, for determining ToAs in the current EPTA project. It is important to note that special cases should not be overlooked, and different combinations should be compared in practical work.

By analyzing the timing RMS variations as a function of the ToA bandwidth and comparing our calculated SLNF with the jitter noise reported in other studies \citep[e.g.][]{pbs+21, lma+19}, we can estimate the current SLNF for various telescope systems and pulsars. It is worth mentioning that the results on the EFF in this paper are all worst-case estimates, and proper calibration is expected to significantly improve these results, although this improvement may be significantly pulsar-dependent. Given that the jitter levels observed in previous research are notably lower than the SLNF, we can conclude that pulse phase jitter noise is not a limiting factor for current EPTA timing accuracy. 

\begin{acknowledgements}
The European Pulsar Timing Array (EPTA) is a collaboration between European and partner institutes, namely ASTRON (NL), INAF/Osservatorio di Cagliari (IT), Max-Planck-Institut f\"ur Radioastronomie (GER), Nan\c{c}ay/Paris Observatory (FRA), the University of Manchester (UK), the University of Birmingham (UK), the University of East Anglia (UK), the University of Bielefeld (GER), the University of Paris (FRA), the University of Milan-Bicocca (IT) and Peking University (CHN), with the aim to provide high precision pulsar timing to work towards the direct detection of low-frequency gravitational waves. An Advanced Grant of the European Research Council to implement the Large European Array for Pulsars (LEAP) also provides funding. J.P.W.V.\  acknowledges support from NSF AccelNet award No. 2114721.

Part of this work is based on observations with the 100-m telescope of the Max-Planck-Institut f\"ur Radioastronomie (MPIfR) at Effelsberg in Germany. Pulsar research at the Jodrell Bank Centre for Astrophysics and the observations using the Lovell Telescope is supported by a Consolidated Grant (ST/T000414/1) from the UK’s Science and Technology Facilities Council. The Nan\c{c}ay Radio Observatory is operated by the Paris Observatory, associated with the French Centre National de la Recherche Scientifique (CNRS), and partially supported by the Region Centre in France. The Westerbork Synthesis Radio Telescope is operated by the Netherlands Institute for Radio Astronomy (ASTRON) with support from The Netherlands Foundation for Scientific Research (NWO).

\end{acknowledgements}


\bibliographystyle{aa}
\bibliography{./Refs/journals,./Refs/psrrefs,./Refs/modrefs,./Refs/crossrefs} % if your bibtex file is called example.bib

\begin{thebibliography}{68}
\expandafter\ifx\csname natexlab\endcsname\relax\def\natexlab#1{#1}\fi

\bibitem[{{Arzoumanian} {et~al.}(2018){Arzoumanian}, {Brazier},
  {Burke-Spolaor}, {Chamberlin}, {Chatterjee}, {Christy}, {Cordes}, {Cornish},
  {Crawford}, {Thankful Cromartie}, {Crowter}, {DeCesar}, {Demorest}, {Dolch},
  {Ellis}, {Ferdman}, {Ferrara}, {Fonseca}, {Garver-Daniels}, {Gentile},
  {Halmrast}, {Huerta}, {Jenet}, {Jessup}, {Jones}, {Jones}, {Kaplan}, {Lam},
  {Lazio}, {Levin}, {Lommen}, {Lorimer}, {Luo}, {Lynch}, {Madison}, {Matthews},
  {McLaughlin}, {McWilliams}, {Mingarelli}, {Ng}, {Nice}, {Pennucci}, {Ransom},
  {Ray}, {Siemens}, {Simon}, {Spiewak}, {Stairs}, {Stinebring}, {Stovall},
  {Swiggum}, {Taylor}, {Vallisneri}, {van Haasteren}, {Vigeland}, {Zhu}, \&
  {NANOGrav Collaboration}}]{abb+18}
{Arzoumanian}, Z., {Brazier}, A., {Burke-Spolaor}, S., {et~al.} 2018, ApJS,
  235, 37

\bibitem[{{Babak} {et~al.}(2016){Babak}, {Petiteau}, {Sesana}, {Brem},
  {Rosado}, {Taylor}, {Lassus}, {Hessels}, {Bassa}, {Burgay}, {Caballero},
  {Champion}, {Cognard}, {Desvignes}, {Gair}, {Guillemot}, {Janssen},
  {Karuppusamy}, {Kramer}, {Lazarus}, {Lee}, {Lentati}, {Liu}, {Mingarelli},
  {Os{\l}owski}, {Perrodin}, {Possenti}, {Purver}, {Sanidas}, {Smits},
  {Stappers}, {Theureau}, {Tiburzi}, {van Haasteren}, {Vecchio}, \&
  {Verbiest}}]{bps+16}
{Babak}, S., {Petiteau}, A., {Sesana}, A., {et~al.} 2016, MNRAS, 455, 1665

\bibitem[{Backer {et~al.}(1982)Backer, Kulkarni, Heiles, Davis, \&
  Goss}]{bkh+82}
Backer, D.~C., Kulkarni, S.~R., Heiles, C., Davis, M.~M., \& Goss, W.~M. 1982,
  Nature, 300, 615

\bibitem[{Bailes {et~al.}(1997)Bailes, Johnston, Bell, Lorimer, Stappers,
  Manchester, Lyne, D'Amico, \& Gaensler}]{bjb+97}
Bailes, M., Johnston, S., Bell, J.~F., {et~al.} 1997, ApJ, 481, 386

\bibitem[{{Bassa} {et~al.}(2016){Bassa}, {Janssen}, {Karuppusamy}, {Kramer},
  {Lee}, {Liu}, {McKee}, {Perrodin}, {Purver}, {Sanidas}, {Smits}, \&
  {Stappers}}]{bjk+16}
{Bassa}, C.~G., {Janssen}, G.~H., {Karuppusamy}, R., {et~al.} 2016, MNRAS, 456,
  2196

\bibitem[{{Burke-Spolaor} {et~al.}(2019){Burke-Spolaor}, {Taylor}, {Charisi},
  {Dolch}, {Hazboun}, {Holgado}, {Kelley}, {Lazio}, {Madison}, {McMann},
  {Mingarelli}, {Rasskazov}, {Siemens}, {Simon}, \& {Smith}}]{btc+19}
{Burke-Spolaor}, S., {Taylor}, S.~R., {Charisi}, M., {et~al.} 2019, Astron.\
  Astropys.\ Rev., 27, 5

\bibitem[{Callanan {et~al.}(1998)Callanan, Garnavich, \& Koester}]{cgk98}
Callanan, P.~J., Garnavich, P.~M., \& Koester, D. 1998, MNRAS, 298, 207

\bibitem[{Camilo {et~al.}(1996)Camilo, Nice, Shrauner, \& Taylor}]{cnst96}
Camilo, F., Nice, D.~J., Shrauner, J.~A., \& Taylor, J.~H. 1996, ApJ, 469, 819

\bibitem[{Camilo {et~al.}(1993)Camilo, Nice, \& Taylor}]{cnt93}
Camilo, F., Nice, D.~J., \& Taylor, J.~H. 1993, ApJ, 412, L37

\bibitem[{{Chalumeau} {et~al.}(2022){Chalumeau}, {Babak}, {Petiteau}, {Chen},
  {Samajdar}, {Caballero}, {Theureau}, {Guillemot}, {Desvignes},
  {Parthasarathy}, {Liu}, {Shaifullah}, {Hu}, {van der Wateren}, {Antoniadis},
  {Bak Nielsen}, {Bassa}, {Berthereau}, {Burgay}, {Champion}, {Cognard},
  {Falxa}, {Ferdman}, {Freire}, {Gair}, {Graikou}, {Guo}, {Jang}, {Janssen},
  {Karuppusamy}, {Keith}, {Kramer}, {Lee}, {Liu}, {Lyne}, {Main}, {McKee},
  {Mickaliger}, {Perera}, {Perrodin}, {Porayko}, {Possenti}, {Sanidas},
  {Sesana}, {Speri}, {Stappers}, {Tiburzi}, {Vecchio}, {Verbiest}, {Wang},
  {Wang}, \& {Xu}}]{cbp+22}
{Chalumeau}, A., {Babak}, S., {Petiteau}, A., {et~al.} 2022, MNRAS, 509, 5538

\bibitem[{{Cognard} {et~al.}(2013){Cognard}, {Theureau}, {Guillemot}, {Liu},
  {Lassus}, \& {Desvignes}}]{ctg+13}
{Cognard}, I., {Theureau}, G., {Guillemot}, L., {et~al.} 2013, in SF2A-2013:
  Proceedings of the Annual meeting of the French Society of Astronomy and
  Astrophysics, ed. L.~{Cambresy}, F.~{Martins}, E.~{Nuss}, \& A.~{Palacios},
  327--330

\bibitem[{{Desvignes} {et~al.}(2016){Desvignes}, {Caballero}, {Lentati},
  {Verbiest}, {Champion}, {Stappers}, {Janssen}, {Lazarus}, {Os{\l}owski},
  {Babak}, {Bassa}, {Brem}, {Burgay}, {Cognard}, {Gair}, {Graikou},
  {Guillemot}, {Hessels}, {Jessner}, {Jordan}, {Karuppusamy}, {Kramer},
  {Lassus}, {Lazaridis}, {Lee}, {Liu}, {Lyne}, {McKee}, {Mingarelli},
  {Perrodin}, {Petiteau}, {Possenti}, {Purver}, {Rosado}, {Sanidas}, {Sesana},
  {Shaifullah}, {Smits}, {Taylor}, {Theureau}, {Tiburzi}, {van Haasteren}, \&
  {Vecchio}}]{dcl+16}
{Desvignes}, G., {Caballero}, R.~N., {Lentati}, L., {et~al.} 2016, MNRAS, 458,
  3341

\bibitem[{{Donner} {et~al.}(2020){Donner}, {Verbiest}, {Tiburzi},
  {Os{\l}owski}, {K{\"u}nsem{\"o}ller}, {Bak Nielsen}, {Grie{\ss}meier},
  {Serylak}, {Kramer}, {Anderson}, {Wucknitz}, {Keane}, {Kondratiev}, {Sobey},
  {McKee}, {Bilous}, {Breton}, {Br{\"u}ggen}, {Ciardi}, {Hoeft}, {van Leeuwen},
  \& {Vocks}}]{dvt+20}
{Donner}, J.~Y., {Verbiest}, J.~P.~W., {Tiburzi}, C., {et~al.} 2020, A\&A, 644,
  A153

\bibitem[{Edwards \& Bailes(2001{\natexlab{a}})}]{eb01}
Edwards, R.~T. \& Bailes, M. 2001{\natexlab{a}}, ApJ, 547, L37

\bibitem[{Edwards \& Bailes(2001{\natexlab{b}})}]{eb01b}
Edwards, R.~T. \& Bailes, M. 2001{\natexlab{b}}, ApJ, 553, 801

\bibitem[{{EPTA Collaboration} {et~al.}(2023){EPTA Collaboration},
  {Antoniadis}, {Babak}, {Bak Nielsen}, {Bassa}, {Berthereau}, {Bonetti},
  {Bortolas}, {Brook}, {Burgay}, {Caballero}, {Chalumeau}, {Champion},
  {Chanlaridis}, {Chen}, {Cognard}, {Desvignes}, {Falxa}, {Ferdman},
  {Franchini}, {Gair}, {Goncharov}, {Graikou}, {Grie{\ss}meier}, {Guillemot},
  {Guo}, {Hu}, {Iraci}, {Izquierdo-Villalba}, {Jang}, {Jawor}, {Janssen},
  {Jessner}, {Karuppusamy}, {Keane}, {Keith}, {Kramer}, {Krishnakumar},
  {Lackeos}, {Lee}, {Liu}, {Liu}, {Lyne}, {McKee}, {Main}, {Mickaliger},
  {Ni{\c{t}}u}, {Parthasarathy}, {Perera}, {Perrodin}, {Petiteau}, {Porayko},
  {Possenti}, {Quelquejay Leclere}, {Samajdar}, {Sanidas}, {Sesana},
  {Shaifullah}, {Speri}, {Spiewak}, {Stappers}, {Susarla}, {Theureau},
  {Tiburzi}, {van der Wateren}, {Vecchio}, {Venkatraman Krishnan}, {Verbiest},
  {Wang}, {Wang}, \& {Wu}}]{eab+23}
{EPTA Collaboration}, {Antoniadis}, J., {Babak}, S., {et~al.} 2023, \aap, 678,
  A48

\bibitem[{{Fonseca} {et~al.}(2016){Fonseca}, {Pennucci}, {Ellis}, {Stairs},
  {Nice}, {Ransom}, {Demorest}, {Arzoumanian}, {Crowter}, {Dolch}, {Ferdman},
  {Gonzalez}, {Jones}, {Jones}, {Lam}, {Levin}, {McLaughlin}, {Stovall},
  {Swiggum}, \& {Zhu}}]{fpe+16}
{Fonseca}, E., {Pennucci}, T.~T., {Ellis}, J.~A., {et~al.} 2016, ApJ, 832, 167

\bibitem[{Foster {et~al.}(1995)Foster, Cadwell, Wolszczan, \&
  Anderson}]{fcwa95}
Foster, R.~S., Cadwell, B.~J., Wolszczan, A., \& Anderson, S.~B. 1995, ApJ,
  454, 826

\bibitem[{{Guillemot} {et~al.}(2023){Guillemot}, {Cognard}, {van Straten},
  {Theureau}, \& {G{\'e}rard}}]{gcv+23}
{Guillemot}, L., {Cognard}, I., {van Straten}, W., {Theureau}, G., \&
  {G{\'e}rard}, E. 2023, \aap, 678, A79

\bibitem[{{Hellings} \& {Downs}(1983)}]{hd83}
{Hellings}, R.~W. \& {Downs}, G.~S. 1983, \apjl, 265, L39

\bibitem[{{Hobbs} {et~al.}(2012){Hobbs}, {Coles}, {Manchester}, {Keith},
  {Shannon}, {Chen}, {Bailes}, {Bhat}, {Burke-Spolaor}, {Champion},
  {Chaudhary}, {Hotan}, {Khoo}, {Kocz}, {Levin}, {Oslowski}, {Preisig}, {Ravi},
  {Reynolds}, {Sarkissian}, {van Straten}, {Verbiest}, {Yardley}, \&
  {You}}]{hcm+12}
{Hobbs}, G., {Coles}, W., {Manchester}, R.~N., {et~al.} 2012, MNRAS, 427, 2780

\bibitem[{{Hobbs} {et~al.}(2020){Hobbs}, {Guo}, {Caballero}, {Coles}, {Lee},
  {Manchester}, {Reardon}, {Matsakis}, {Tong}, {Arzoumanian}, {Bailes},
  {Bassa}, {Bhat}, {Brazier}, {Burke-Spolaor}, {Champion}, {Chatterjee},
  {Cognard}, {Dai}, {Desvignes}, {Dolch}, {Ferdman}, {Graikou}, {Guillemot},
  {Janssen}, {Keith}, {Kerr}, {Kramer}, {Lam}, {Liu}, {Lyne}, {Lazio}, {Lynch},
  {McKee}, {McLaughlin}, {Mingarelli}, {Nice}, {Os{\l}owski}, {Pennucci},
  {Perera}, {Perrodin}, {Possenti}, {Russell}, {Sanidas}, {Sesana},
  {Shaifullah}, {Shannon}, {Simon}, {Spiewak}, {Stairs}, {Stappers}, {Swiggum},
  {Taylor}, {Theureau}, {Toomey}, {van Haasteren}, {Wang}, {Wang}, \&
  {Zhu}}]{hgc+20}
{Hobbs}, G., {Guo}, L., {Caballero}, R.~N., {et~al.} 2020, MNRAS, 491, 5951

\bibitem[{{Hotan} {et~al.}(2004{\natexlab{a}}){Hotan}, {Bailes}, \&
  {Ord}}]{hbo04}
{Hotan}, A.~W., {Bailes}, M., \& {Ord}, S.~M. 2004{\natexlab{a}}, MNRAS, 355,
  941

\bibitem[{{Hotan} {et~al.}(2005){Hotan}, {Bailes}, \& {Ord}}]{hbo05a}
{Hotan}, A.~W., {Bailes}, M., \& {Ord}, S.~M. 2005, MNRAS, 362, 1267

\bibitem[{{Hotan} {et~al.}(2004{\natexlab{b}}){Hotan}, {van Straten}, \&
  {Manchester}}]{hvm04}
{Hotan}, A.~W., {van Straten}, W., \& {Manchester}, R.~N. 2004{\natexlab{b}},
  PASA, 21, 302

\bibitem[{Jacoby {et~al.}(2007)Jacoby, Bailes, Ord, Knight, \& Hotan}]{jbo+07}
Jacoby, B.~A., Bailes, M., Ord, S.~M., Knight, H.~S., \& Hotan, A.~W. 2007,
  ApJ, 656, 408

\bibitem[{{Jacoby} {et~al.}(2003){Jacoby}, {Bailes}, {van Kerkwijk}, {Ord},
  {Hotan}, {Kulkarni}, \& {Anderson}}]{jbv+03}
{Jacoby}, B.~A., {Bailes}, M., {van Kerkwijk}, M.~H., {et~al.} 2003, ApJ, 599,
  L99

\bibitem[{{Janssen} {et~al.}(2010){Janssen}, {Stappers}, {Bassa}, {Cognard},
  {Kramer}, \& {Theureau}}]{jsb+10}
{Janssen}, G.~H., {Stappers}, B.~W., {Bassa}, C.~G., {et~al.} 2010, A\&A, 514,
  A74+

\bibitem[{{Karuppusamy} {et~al.}(2008){Karuppusamy}, {Stappers}, \& {van
  Straten}}]{ksv08}
{Karuppusamy}, R., {Stappers}, B., \& {van Straten}, W. 2008, PASP, 120, 191

\bibitem[{{Kramer} {et~al.}(2021){Kramer}, {Stairs}, {Manchester}, {Wex},
  {Deller}, {Coles}, {Ali}, {Burgay}, {Camilo}, {Cognard}, {Damour},
  {Desvignes}, {Ferdman}, {Freire}, {Grondin}, {Guillemot}, {Hobbs}, {Janssen},
  {Karuppusamy}, {Lorimer}, {Lyne}, {McKee}, {McLaughlin}, {M{\"u}nch},
  {Perera}, {Pol}, {Possenti}, {Sarkissian}, {Stappers}, \&
  {Theureau}}]{ksm+21}
{Kramer}, M., {Stairs}, I.~H., {Manchester}, R.~N., {et~al.} 2021, Physical
  Review X, 11, 041050

\bibitem[{Kramer {et~al.}(1999)Kramer, Xilouris, Camilo, Nice, Lange, Backer,
  \& Doroshenko}]{kxc+99}
Kramer, M., Xilouris, K.~M., Camilo, F., {et~al.} 1999, ApJ, 520, 324

\bibitem[{{Lam} {et~al.}(2019){Lam}, {McLaughlin}, {Arzoumanian}, {Blumer},
  {Brook}, {Cromartie}, {Demorest}, {DeCesar}, {Dolch}, {Ellis}, {Ferdman},
  {Ferrara}, {Fonseca}, {Garver-Daniels}, {Gentile}, {Jones}, {Lorimer},
  {Lynch}, {Ng}, {Nice}, {Pennucci}, {Ransom}, {Spiewak}, {Stairs}, {Stovall},
  {Swiggum}, {Vigeland}, \& {Zhu}}]{lma+19}
{Lam}, M.~T., {McLaughlin}, M.~A., {Arzoumanian}, Z., {et~al.} 2019, ApJ, 872,
  193

\bibitem[{Lange {et~al.}(2001)Lange, Camilo, Wex, Kramer, Backer, Lyne, \&
  Doroshenko}]{lcw+01}
Lange, C., Camilo, F., Wex, N., {et~al.} 2001, MNRAS, 326, 274

\bibitem[{{Lattimer} \& {Prakash}(2016)}]{lp16}
{Lattimer}, J.~M. \& {Prakash}, M. 2016, Phys. Rep., 621, 127

\bibitem[{Lazaridis {et~al.}(2009)Lazaridis, Wex, Jessner, Kramer, Stappers,
  Janssen, Desvignes, Purver, Cognard, Theureau, Lyne, Jordan, \&
  Zensus}]{lwj+09}
Lazaridis, K., Wex, N., Jessner, A., {et~al.} 2009, MNRAS, 400, 805

\bibitem[{{Lazarus} {et~al.}(2016){Lazarus}, {Karuppusamy}, {Graikou},
  {Caballero}, {Champion}, {Lee}, {Verbiest}, \& {Kramer}}]{lkg+16}
{Lazarus}, P., {Karuppusamy}, R., {Graikou}, E., {et~al.} 2016, MNRAS, 458, 868

\bibitem[{{Lentati} \& {Shannon}(2015)}]{ls15}
{Lentati}, L. \& {Shannon}, R.~M. 2015, MNRAS, 454, 1058

\bibitem[{{Liu} {et~al.}(2014){Liu}, {Desvignes}, {Cognard}, {Stappers},
  {Verbiest}, {Lee}, {Champion}, {Kramer}, {Freire}, \& {Karuppusamy}}]{ldc+14}
{Liu}, K., {Desvignes}, G., {Cognard}, I., {et~al.} 2014, MNRAS, 443, 3752

\bibitem[{{Liu} {et~al.}(2020){Liu}, {Guillemot}, {Istrate}, {Shao}, {Tauris},
  {Wex}, {Antoniadis}, {Chalumeau}, {Cognard}, {Desvignes}, {Freire}, {Kehl},
  \& {Theureau}}]{lgi+20}
{Liu}, K., {Guillemot}, L., {Istrate}, A.~G., {et~al.} 2020, MNRAS, 499, 2276

\bibitem[{{Liu} {et~al.}(2015){Liu}, {Karuppusamy}, {Lee}, {Stappers},
  {Kramer}, {Smits}, {Purver}, {Janssen}, \& {Perrodin}}]{lkl+15}
{Liu}, K., {Karuppusamy}, R., {Lee}, K.~J., {et~al.} 2015, MNRAS, 449, 1158

\bibitem[{{Liu} {et~al.}(2022){Liu}, {Verbiest}, {Main}, {Wu}, {Ambalappat},
  {Champion}, {Cognard}, {Guillemot}, {Gaikwad}, {Janssen}, {Kramer}, {Keith},
  {Karuppusamy}, {K{\"u}nkel}, {Liu}, {McKee}, {Mickaliger}, {Stappers},
  {Shaifullah}, \& {Theureau}}]{lvm+22}
{Liu}, Y., {Verbiest}, J. P.~W., {Main}, R.~A., {et~al.} 2022, \aap, 664, A116

\bibitem[{{Lommen} {et~al.}(2000){Lommen}, {Zepka}, {Backer}, {McLaughlin},
  {Cordes}, {Arzoumanian}, \& {Xilouris}}]{lzb+00}
{Lommen}, A.~N., {Zepka}, A., {Backer}, D.~C., {et~al.} 2000, ApJ, 545, 1007

\bibitem[{Lorimer {et~al.}(1995)Lorimer, Nicastro, Lyne, Bailes, Manchester,
  Johnston, Bell, D'Amico, \& Harrison}]{lnl+95}
Lorimer, D.~R., Nicastro, L., Lyne, A.~G., {et~al.} 1995, ApJ, 439, 933

\bibitem[{{Manchester} {et~al.}(2013){Manchester}, {Hobbs}, {Bailes}, {Coles},
  {van Straten}, {Keith}, {Shannon}, {Bhat}, {Brown}, {Burke-Spolaor},
  {Champion}, {Chaudhary}, {Edwards}, {Hampson}, {Hotan}, {Jameson}, {Jenet},
  {Kesteven}, {Khoo}, {Kocz}, {Maciesiak}, {Oslowski}, {Ravi}, {Reynolds},
  {Sarkissian}, {Verbiest}, {Wen}, {Wilson}, {Yardley}, {Yan}, \&
  {You}}]{mhb+13}
{Manchester}, R.~N., {Hobbs}, G., {Bailes}, M., {et~al.} 2013, PASA, 30, 17

\bibitem[{Misner {et~al.}(1973)Misner, Thorne, \& Wheeler}]{mtw73}
Misner, C.~W., Thorne, K.~S., \& Wheeler, J.~A. 1973, Gravitation (San
  Francisco: W. H. Freeman)

\bibitem[{{Morello} {et~al.}(2019){Morello}, {Barr}, {Cooper}, {Bailes},
  {Bates}, {Bhat}, {Burgay}, {Burke-Spolaor}, {Cameron}, {Champion}, {Eatough},
  {Flynn}, {Jameson}, {Johnston}, {Keith}, {Keane}, {Kramer}, {Levin}, {Ng},
  {Petroff}, {Possenti}, {Stappers}, {van Straten}, \& {Tiburzi}}]{mbc+19}
{Morello}, V., {Barr}, E.~D., {Cooper}, S., {et~al.} 2019, MNRAS, 483, 3673

\bibitem[{Nicastro {et~al.}(1995)Nicastro, Lyne, Lorimer, Harrison, Bailes, \&
  Skidmore}]{nll+95}
Nicastro, L., Lyne, A.~G., Lorimer, D.~R., {et~al.} 1995, MNRAS, 273, L68

\bibitem[{Ord {et~al.}(2006)Ord, Jacoby, Hotan, \& Bailes}]{ojhb06}
Ord, S.~M., Jacoby, B.~A., Hotan, A.~W., \& Bailes, M. 2006, MNRAS, 371, 337

\bibitem[{{{\"O}zel} \& {Freire}(2016)}]{of16}
{{\"O}zel}, F. \& {Freire}, P. 2016, Ann. Rev. Astr. Ap., 54, 401

\bibitem[{{Parthasarathy} {et~al.}(2021){Parthasarathy}, {Bailes}, {Shannon},
  {van Straten}, {Os{\l}owski}, {Johnston}, {Spiewak}, {Reardon}, {Kramer},
  {Venkatraman Krishnan}, {Pennucci}, {Abbate}, {Buchner}, {Camilo},
  {Champion}, {Geyer}, {Hugo}, {Jameson}, {Karastergiou}, {Keith}, \&
  {Serylak}}]{pbs+21}
{Parthasarathy}, A., {Bailes}, M., {Shannon}, R.~M., {et~al.} 2021, MNRAS, 502,
  407

\bibitem[{{Perera} {et~al.}(2019){Perera}, {DeCesar}, {Demorest}, {Kerr},
  {Lentati}, {Nice}, {Os{\l}owski}, {Ransom}, {Keith}, {Arzoumanian}, {Bailes},
  {Baker}, {Bassa}, {Bhat}, {Brazier}, {Burgay}, {Burke-Spolaor}, {Caballero},
  {Champion}, {Chatterjee}, {Chen}, {Cognard}, {Cordes}, {Crowter}, {Dai},
  {Desvignes}, {Dolch}, {Ferdman}, {Ferrara}, {Fonseca}, {Goldstein},
  {Graikou}, {Guillemot}, {Hazboun}, {Hobbs}, {Hu}, {Islo}, {Janssen},
  {Karuppusamy}, {Kramer}, {Lam}, {Lee}, {Liu}, {Luo}, {Lyne}, {Manchester},
  {McKee}, {McLaughlin}, {Mingarelli}, {Parthasarathy}, {Pennucci}, {Perrodin},
  {Possenti}, {Reardon}, {Russell}, {Sanidas}, {Sesana}, {Shaifullah},
  {Shannon}, {Siemens}, {Simon}, {Spiewak}, {Stairs}, {Stappers}, {Swiggum},
  {Taylor}, {Theureau}, {Tiburzi}, {Vallisneri}, {Vecchio}, {Wang}, {Zhang},
  {Zhang}, {Zhu}, \& {Zhu}}]{pdd+19}
{Perera}, B.~B.~P., {DeCesar}, M.~E., {Demorest}, P.~B., {et~al.} 2019, MNRAS,
  490, 4666

\bibitem[{{Ramachandran} \& {Kramer}(2003)}]{rk03}
{Ramachandran}, R. \& {Kramer}, M. 2003, A\&A, 407, 1085

\bibitem[{Sayer {et~al.}(1997)Sayer, Nice, \& Taylor}]{snt97}
Sayer, R.~W., Nice, D.~J., \& Taylor, J.~H. 1997, ApJ, 474, 426

\bibitem[{Segelstein {et~al.}(1986)Segelstein, Rawley, Stinebring, Fruchter, \&
  Taylor}]{srs+86}
Segelstein, D.~J., Rawley, L.~A., Stinebring, D.~R., Fruchter, A.~S., \&
  Taylor, J.~H. 1986, Nature, 322, 714

\bibitem[{{Shannon} {et~al.}(2014){Shannon}, {Os{\l}owski}, {Dai}, {Bailes},
  {Hobbs}, {Manchester}, {van Straten}, {Raithel}, {Ravi}, {Toomey}, {Bhat},
  {Burke-Spolaor}, {Coles}, {Keith}, {Kerr}, {Levin}, {Sarkissian}, {Wang},
  {Wen}, \& {Zhu}}]{sod+14}
{Shannon}, R.~M., {Os{\l}owski}, S., {Dai}, S., {et~al.} 2014, MNRAS, 443, 1463

\bibitem[{{Stappers} {et~al.}(2006){Stappers}, {Kramer}, {Lyne}, {D'Amico}, \&
  {Jessner}}]{skl+06}
{Stappers}, B.~W., {Kramer}, M., {Lyne}, A.~G., {D'Amico}, N., \& {Jessner}, A.
  2006, Chin.\ J.\ Astron.\ Astrophys., 6, 298

\bibitem[{Tiburzi {et~al.}(2021)Tiburzi, Shaifullah, Bassa, Zucca, Verbiest,
  Porayko, van~der Wateren, Fallows, Main, Janssen, {et~al.}}]{tsb+21}
Tiburzi, C., Shaifullah, G., Bassa, C., {et~al.} 2021, A\&A, 647, A84

\bibitem[{Tiburzi {et~al.}(2019)Tiburzi, Verbiest, Shaifullah, Janssen,
  Anderson, Horneffer, K{\"u}nsem{\"o}ller, Os{\l}owski, Donner, Kramer,
  {et~al.}}]{tvs+19}
Tiburzi, C., Verbiest, J., Shaifullah, G., {et~al.} 2019, MNRAS, 487, 394

\bibitem[{{Tukey}(1977)}]{tuk77}
{Tukey}, J.~W. 1977, {Exploratory data analysis} (Reading, MA)

\bibitem[{{Turner} {et~al.}(2021){Turner}, {McLaughlin}, {Cordes}, {Lam},
  {Shapiro-Albert}, {Stinebring}, {Arzoumanian}, {Blumer}, {Brook},
  {Chatterjee}, {Cromartie}, {DeCesar}, {Demorest}, {Dolch}, {Ellis},
  {Ferdman}, {Ferrara}, {Fonseca}, {Garver-Daniels}, {Gentile}, {Good},
  {Jones}, {Lazio}, {Lorimer}, {Luo}, {Lynch}, {Ng}, {Nice}, {Pennucci}, {Pol},
  {Ransom}, {Spiewak}, {Stairs}, {Stovall}, {Swiggum}, \& {Vigeland}}]{tmc+21}
{Turner}, J.~E., {McLaughlin}, M.~A., {Cordes}, J.~M., {et~al.} 2021, ApJ, 917,
  10

\bibitem[{van Kerkwijk {et~al.}(1996)van Kerkwijk, Bergeron, \&
  Kulkarni}]{vbk96}
van Kerkwijk, M.~H., Bergeron, P., \& Kulkarni, S.~R. 1996, ApJ, 467, L89

\bibitem[{{van Straten}(2013)}]{van13}
{van Straten}, W. 2013, ApJS, 204, 13

\bibitem[{{Verbiest} {et~al.}(2008){Verbiest}, {Bailes}, {van Straten},
  {Hobbs}, {Edwards}, {Manchester}, {Bhat}, {Sarkissian}, {Jacoby}, \&
  {Kulkarni}}]{vbv+08}
{Verbiest}, J.~P.~W., {Bailes}, M., {van Straten}, W., {et~al.} 2008, ApJ, 679,
  675

\bibitem[{{Verbiest} {et~al.}(2016){Verbiest}, {Lentati}, {Hobbs}, {van
  Haasteren}, {Demorest}, {Janssen}, {Wang}, {Desvignes}, {Caballero}, {Keith},
  {Champion}, {Arzoumanian}, {Babak}, {Bassa}, {Bhat}, {Brazier}, {Brem},
  {Burgay}, {Burke-Spolaor}, {Chamberlin}, {Chatterjee}, {Christy}, {Cognard},
  {Cordes}, {Dai}, {Dolch}, {Ellis}, {Ferdman}, {Fonseca}, {Gair},
  {Garver-Daniels}, {Gentile}, {Gonzalez}, {Graikou}, {Guillemot}, {Hessels},
  {Jones}, {Karuppusamy}, {Kerr}, {Kramer}, {Lam}, {Lasky}, {Lassus},
  {Lazarus}, {Lazio}, {Lee}, {Levin}, {Liu}, {Lynch}, {Lyne}, {Mckee},
  {McLaughlin}, {McWilliams}, {Madison}, {Manchester}, {Mingarelli}, {Nice},
  {Os{\l}owski}, {Palliyaguru}, {Pennucci}, {Perera}, {Perrodin}, {Possenti},
  {Petiteau}, {Ransom}, {Reardon}, {Rosado}, {Sanidas}, {Sesana}, {Shaifullah},
  {Shannon}, {Siemens}, {Simon}, {Smits}, {Spiewak}, {Stairs}, {Stappers},
  {Stinebring}, {Stovall}, {Swiggum}, {Taylor}, {Theureau}, {Tiburzi},
  {Toomey}, {Vallisneri}, {van Straten}, {Vecchio}, {Wang}, {Wen}, {You},
  {Zhu}, \& {Zhu}}]{vlh+16}
{Verbiest}, J.~P.~W., {Lentati}, L., {Hobbs}, G., {et~al.} 2016, MNRAS, 458,
  1267

\bibitem[{{Verbiest} \& {Shaifullah}(2018)}]{vs18}
{Verbiest}, J.~P.~W. \& {Shaifullah}, G. 2018, Classical and Quantum Gravity,
  35

\bibitem[{{Wang} {et~al.}(2022){Wang}, {Shaifullah}, {Verbiest}, {Tiburzi},
  {Champion}, {Cognard}, {Gaikwad}, {Graikou}, {Guillemot}, {Hu},
  {Karuppusamy}, {Keith}, {Kramer}, {Liu}, {Lyne}, {Mickaliger}, {Stappers}, \&
  {Theureau}}]{wsv+22}
{Wang}, J., {Shaifullah}, G.~M., {Verbiest}, J.~P.~W., {et~al.} 2022, \aap,
  658, A181

\bibitem[{{Wang} {et~al.}(2023){Wang}, {Verbiest}, {Shaifullah}, \&
  {Yuan}}]{wvsy23}
{Wang}, J., {Verbiest}, J.~P.~W., {Shaifullah}, G.~M., \& {Yuan}, J.~P. 2023,
  RAA, 23, 125020

\bibitem[{{Zhu} {et~al.}(2015){Zhu}, {Stairs}, {Demorest}, {Nice}, {Ellis},
  {Ransom}, {Arzoumanian}, {Crowter}, {Dolch}, {Ferdman}, {Fonseca},
  {Gonzalez}, {Jones}, {Jones}, {Lam}, {Levin}, {McLaughlin}, {Pennucci},
  {Stovall}, \& {Swiggum}}]{zsd+15}
{Zhu}, W.~W., {Stairs}, I.~H., {Demorest}, P.~B., {et~al.} 2015, ApJ, 809, 41

\end{thebibliography}

% \begin{appendix} %First appendix
\appendix
\section{Pulse profiles and profile differences}
\begin{figure*}[!htbp]
    \centering
 \includegraphics[width=0.85\textwidth]{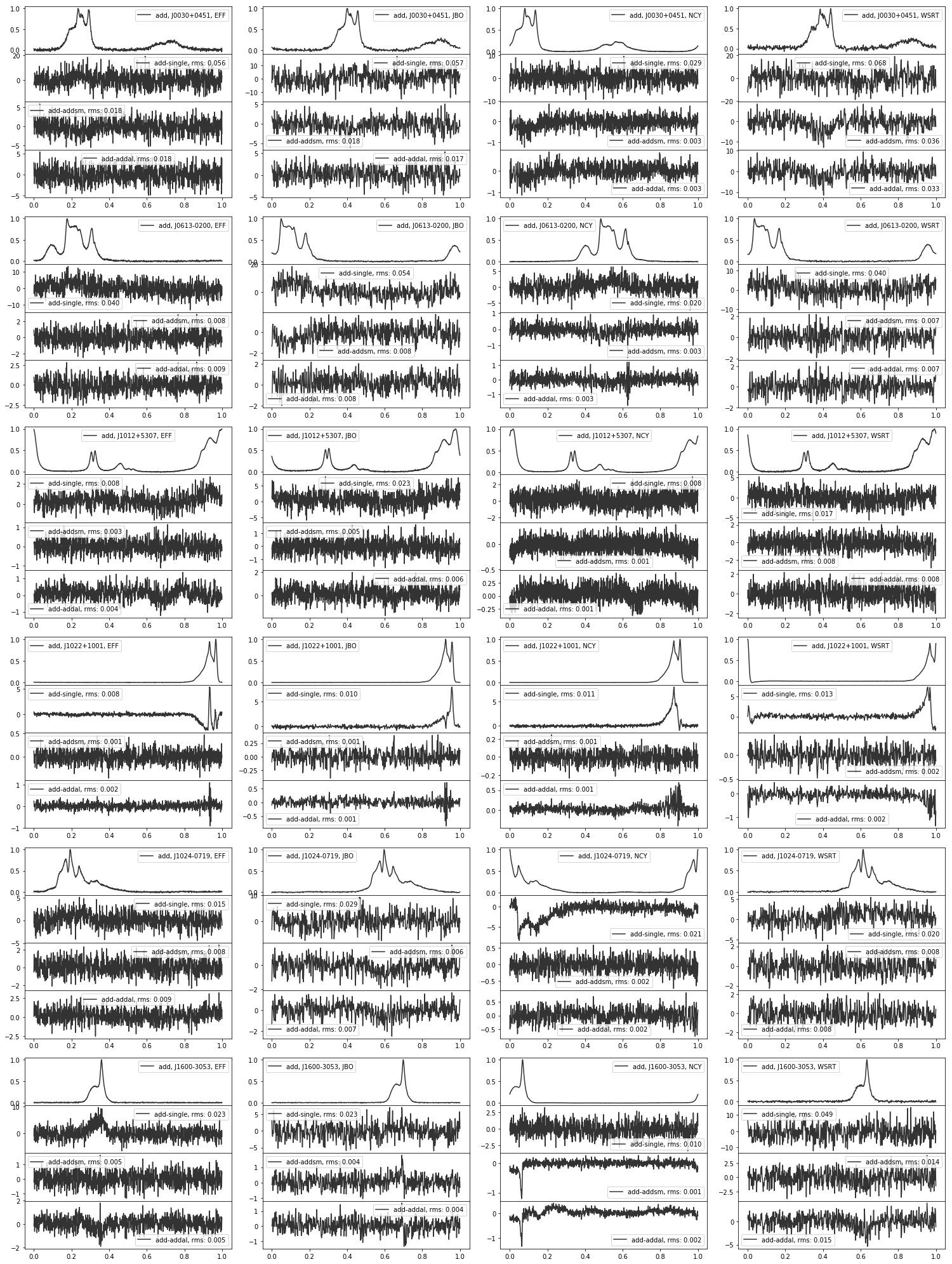}
 \caption[Pulsar profiles and profile differences.]{Pulse profiles of the added templates and profile differences between added and single, analytic, and smoothed templates. From top to bottom, PSRs~J0030+0451, J0613$-$0200, J1012+5307, J1022+1001, J1024$-$0719 and J1600$-$3053. Within each subplot, the profile of each pulsar and the difference between added and single, smoothed and analytic templates are shown from top to bottom respectively. In each row, the EFF, JBO, NCY and WSRT data are shown from left to right.}
 \label{fig:pprof1}
\end{figure*}

\begin{figure*}[!htbp]
\ContinuedFloat
    \centering
 \includegraphics[width=0.95\textwidth]{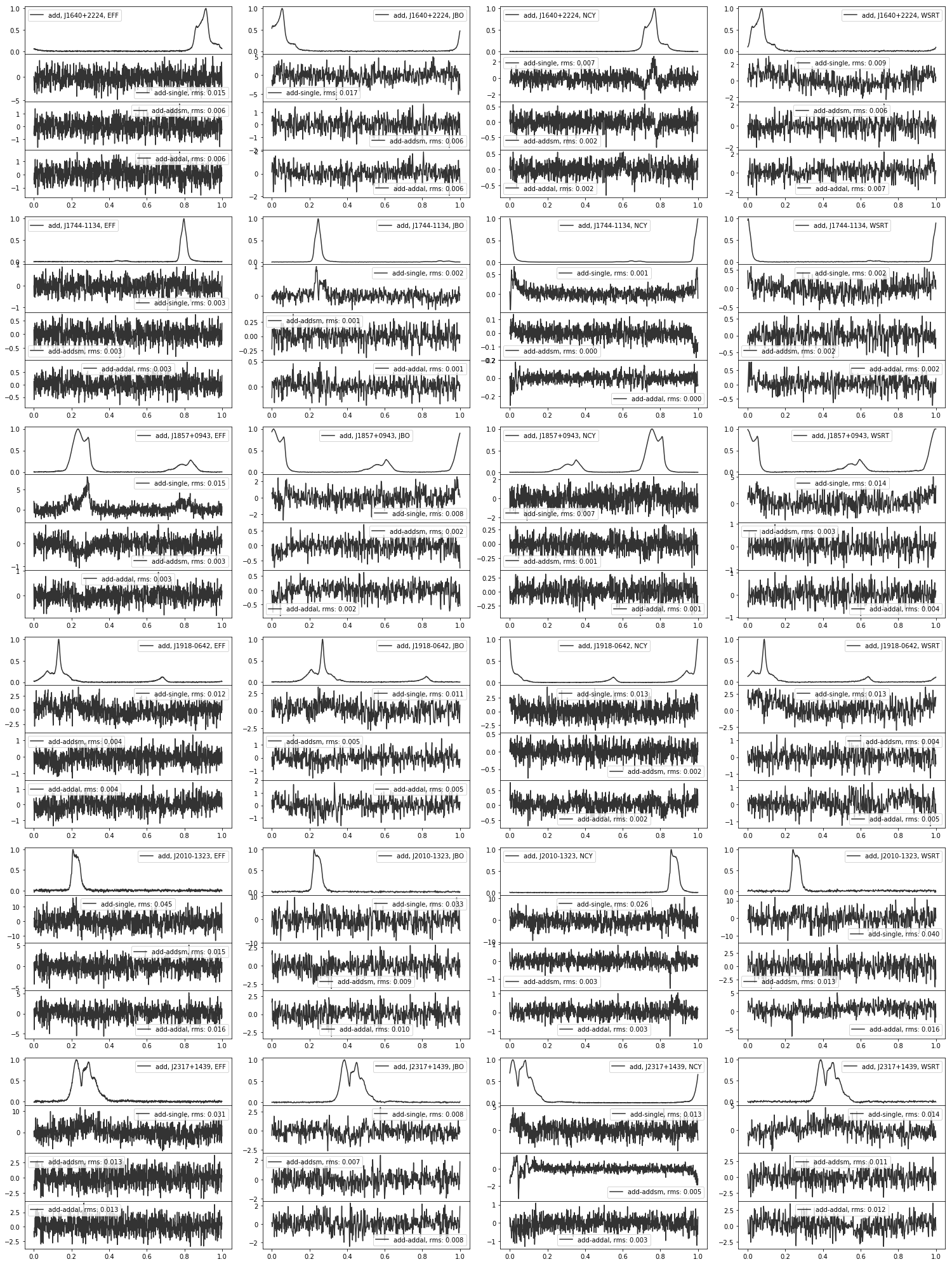}
 \caption{continued for the pulsars PSRs~J1640+2224, J1744$-$1134, J1857+0943, J1918$-$0642, J2010$-$1323, J2317+1439.}
 % \label{fig:pprof2}
\end{figure*}

\section{$\chi_r^2$ and residual RMS}
\begin{figure*}[!htbp]
    \centering
 \includegraphics[width=0.85\textwidth]{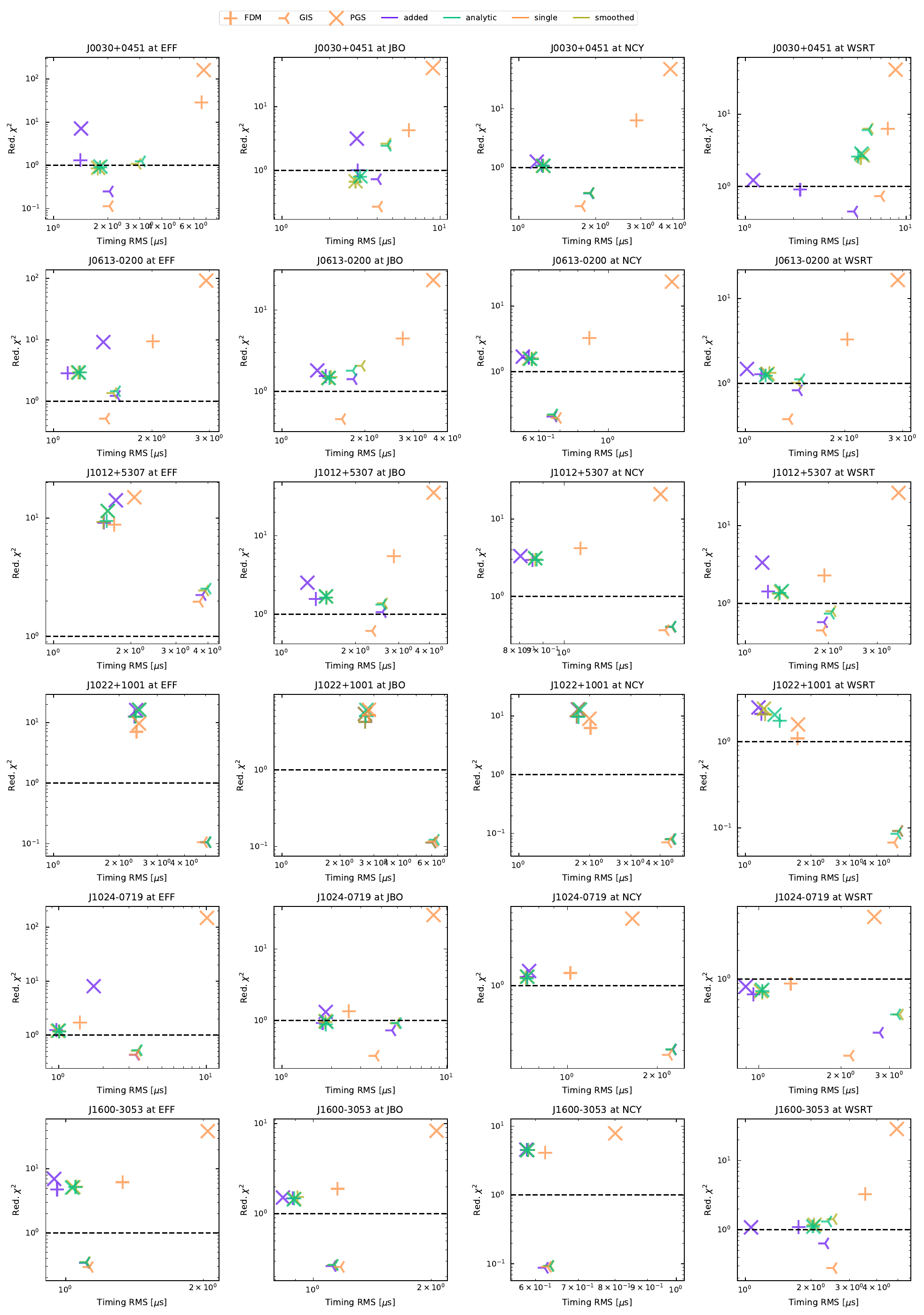}
 \caption[$\chi_r^2$ and residual RMS.]{$\chi_r^2$ and residual RMS for the same pulsars as in the first panel of Fig~\ref{fig:pprof1}, as a function of the selected TMM (marker shape) and template (marker color) and the different telescopes.}
 \label{fig:temp_diff01}
\end{figure*}

\begin{figure*}[!htbp]
\ContinuedFloat
    \centering
 \includegraphics[width=0.95\textwidth]{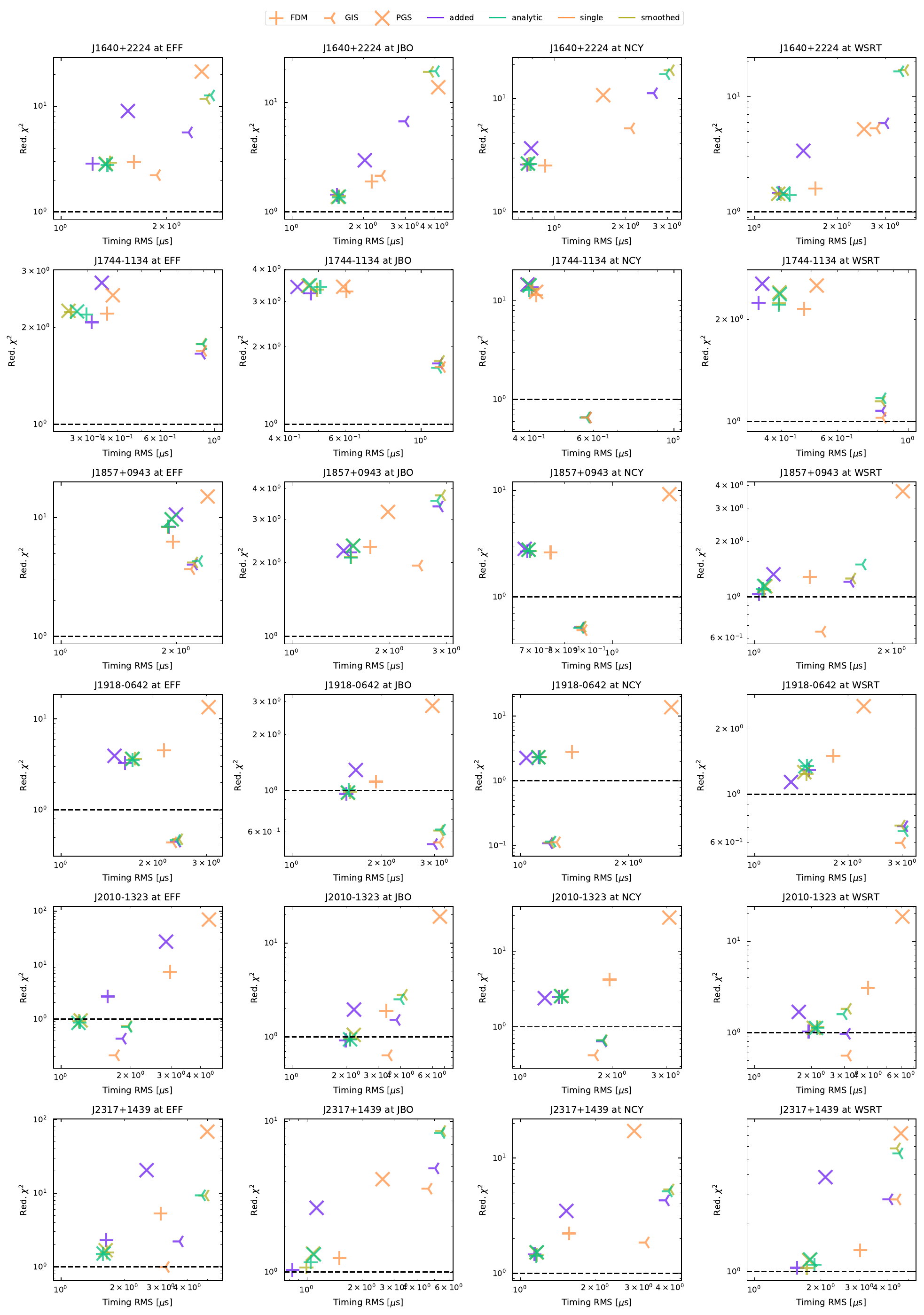}
 \caption{continued for the same pulsars as in the second panel of Fig~\ref{fig:pprof1}}
 % \label{fig:temp_diff02}
\end{figure*}

\section{SLNF estimates for the remaining 12 pulsars}

\begin{figure*}[!htbp]
\centering
\includegraphics[width=0.82\textwidth]{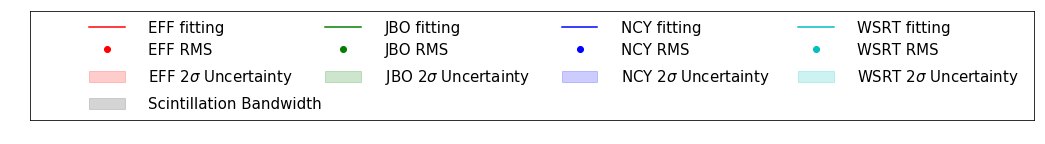}
 \includegraphics[width=0.9\textwidth, scale=0.85]{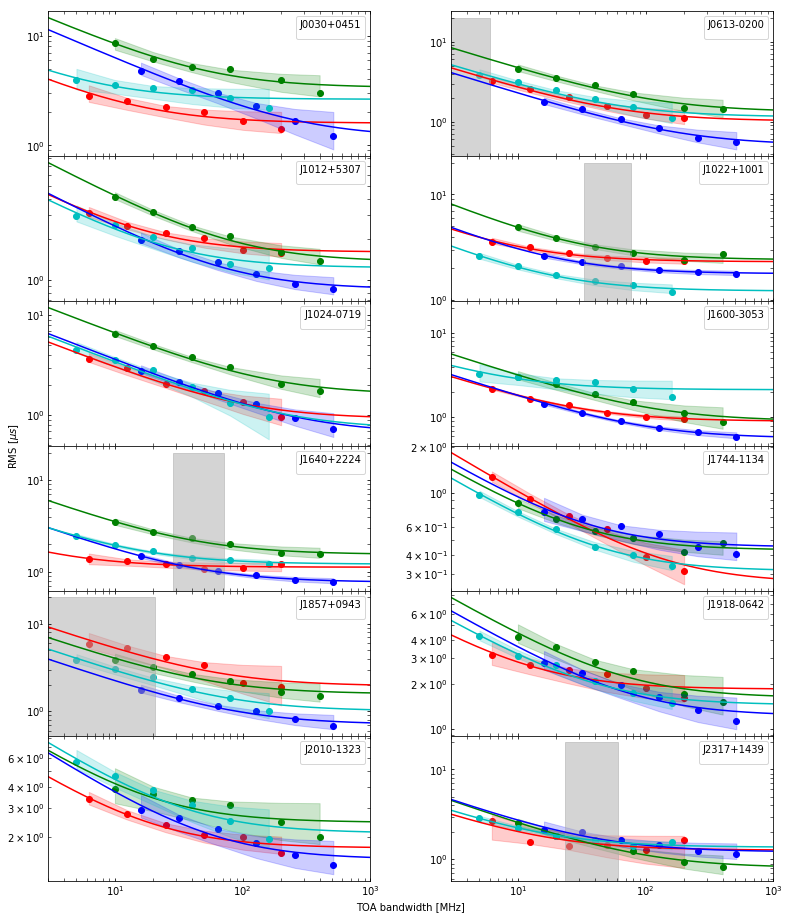}
 \caption{SLNF estimates for the remaining 12 pulsars.}
 \label{fig:psr_toabw}
\end{figure*}
  
% \end{appendix}
%
%
\end{document}